\documentclass[12pt]{article}
\usepackage{amsmath}
\usepackage{amssymb}
\usepackage{epsfig}
\usepackage{cite}
\usepackage{color,colordvi}

\begin{document}
\vspace{0.01cm}
\begin{center}
{\Large\bf   Traces and Time: a de Sitter Black Hole correspondence} 

\end{center}

\vspace{0.1cm}
%\end{center}

\begin{center}

{\bf Cesar Gomez}$^{a}$\footnote{cesar.gomez@uam.es}

\vspace{.6truecm}

{\em $^a$
Instituto de F\'{\i}sica Te\'orica UAM-CSIC\\
Universidad Aut\'onoma de Madrid,
Cantoblanco, 28049 Madrid, Spain}\\

\end{center}

%\vspace{0.5cm}

\begin{abstract}
\noindent  
 
{ We describe how general covariance for QFT defined on a space-time background with horizons leads to the need of adding an extra quantum degree of freedom. The definition of traces and entropies involves the use of a formally thermal state (weight) on the algebra of observables of the added degree of freedom. This extra degree of freedom is promoted into a physical clock after defining an hermitian and positive clock Hamiltonian. The so modified weight can be used to associate to both dS and the BH a type $II_1$ factor and a finite trace. The Murray-von Neumann dimension ( coupling constant) for this type $II_1$ description of the black hole naturally produces an algebraic version of Page's curve.
}

\end{abstract}

\thispagestyle{empty}
\clearpage
\tableofcontents
\newpage
%\section{Introduction}
%\part{Introduction}
\section{Introduction and Summary}

Probably the two main results of quantum gravity are the Bekenstein-Hawking (BH) entropy formula for black holes \cite{Bek}\cite{Hawking} and the analog Gibbons-Hawking (GH) formula for the de Sitter entropy \cite{GH}. In essence both formulas are upper  bounds on the amount of empirical  {\it information} that is {\it inaccessible} to either an external asymptotic observer, in the black hole case, or to a generic dS observer surrounded by her cosmological horizon. These upper bounds have an extremely nice geometrical interpretation as $\frac{A}{4G}$ for $A$ the horizon area either of the black hole or the observer's cosmological horizon. Quantum mechanically we expect for both formulas a simple mathematical representation as
\begin{equation}\label{main}
\ln tr (1)
\end{equation}
where, and here lies the enormous conceptual difficulty, the apparently innocent $1$ should represent {\it the identity for the algebra of physical observables empirically available to the observer} and where $tr$ {\it should be defined by some well defined linear form on this algebra  satisfying the trace property}. In other words (\ref{main}) requires to identify both, the algebra of observables as well as the quantum Hilbert space on which this algebra is  acting. To identify both components entering into (\ref{main}) is what we normally understand as a {\it microscopic model of either the BH entropy for black holes or the GH entropy for de Sitter}. Underlying the basic mathematical difficulties implicit in the microscopic representation (\ref{main}) is the well known fact that traces ( of identities) are counting the dimension of the relevant Hilbert space of states. In quantum mechanics, and for systems admitting a space-time interpretation, this dimension is, irrespectively how small is the number of degrees of freedom, infinity. The finiteness, in quantum gravity, of BH or GH entropy formulas i.e. $\frac{A}{4G}$, has motivated a holographic \cite{thooft, Susskind1} {\it quantum information} representation of the microscopic Hilbert space with $tr(1)$ representing the effective number of q-bits needed to account for the total amount of inaccessible information. The recent approach based on von Neumann algebras \cite{LL1,LL2,Witten1,Witten2,Gomez1,Witten3,Witten4,Witten5,Gomez2,Corean,LL3,Gomez3,recent1,recent2,recent3} addresses the problem of identifying, in the weak gravity limit, the algebras, as well as the Hilbert space of states on which they are acting, describing the limited QFT knowledge of observers living in {\it fixed space-times backgrounds} with horizons.

An important lesson of the mathematical development of the theory of von Neumann algebras \footnote{The reader can find most of the mathematical material used in this note in \cite{Jones}.}, basically due to the work of Tomita,Takesaki and Connes, among others, is to unveil a deep relation between traces and {\it time}. In this note we will discuss how to extend this connection when we replace {\it time} by its empirical manifestation in the form of {\it physical clocks}. 

To identify the frame where we will be working we summarize in the rest of this introduction the basic elements of our further discussion. 

\vspace{0.9cm}

In local quantum field theory (QFT) defined on a fixed space-time background and in the weak gravity limit where we reduce the quantum fluctuations of gravitational origin to those defining linearized gravity on the fixed background, we associate with a generic observer \footnote{Defined by a timelike world line extending from time $-\infty$ to $+\infty$.} the corresponding causal domain ${\cal{D}}$ and the algebra ${\cal{A}}({\cal{D}})$ of QFT local observables with support in ${\cal{D}}$. We normally require that the algebra ${\cal{A}}({\cal{D}})$ is closed under the action of the {\it adjoint operation} and that it contains an {\it identity}. However when we do QFT we {\it don't define the algebra in purely abstract algebraic terms as a $C^{*}$ algebra} but as an  algebra of bounded operators {\it acting on some concrete Hilbert space $H_{QFT}$}. In other words, as a {\it subalgebra} of the algebra $B(H_{QFT})$ of bounded operators acting on $H_{QFT}$. 

For some space-time backgrounds, as Minkowski, the causal domain of a generic observer covers the whole space-time. This is not the case when we consider more general space-time backgrounds with {\it horizons}. In the case of positive cosmological constant, the de Sitter causal domain of a generic observer is {\it the static patch} limited by the corresponding cosmological horizon. Thus, in this case, the algebra ${\cal{A}}({\cal{D}})$ only contains observable information about the static patch. The same situation takes place for the asymptotic observer in the space-time background defined by the eternal black hole. In that case the asymptotic observer algebra ${\cal{A}}({\cal{D}})$ only contains observable information about the black hole exterior. 

Something specially important in the definition of ${\cal{A}}({\cal{D}})$ is that the algebra is defined as {\it acting} on the {\it full Hilbert space} $H_{QFT}$. States in $H_{QFT}$ describe, in principle, the {\it full space-time} not only the causal domain ${\cal{D}}$. In this sense the definition of the Hilbert space $H_{QFT}$ goes beyond the {\it empirical information} available to the observer. One important aspect of $H_{QFT}$ is that we can use it to define a representation of the full group $G$ of {\it isometries} of the space-time background. Moreover, we can identify in $H_{QFT}$ a special state invariant under the whole group $G$ and to think in this state as the quantum representative of the empty fixed background geometry. Note that in general the group $G$ is larger than the group of isometries preserving the observer causal domain ${\cal{D}}$.

In order to make contact with experiments the observer will need some way to associate with any self adjoint observable $a$ in ${\cal{A}}({\cal{D}})$ its {\it expectation value}. To do that you need to define a {\it linear form} $\phi$ defined on ${\cal{A}}({\cal{D}})$ with $\phi(a)$ encoding the expectation value we expect to obtain performing measurements of $a$ in the causal domain. Obviously we require $\phi$ to be linear to account for the quantum mechanical superposition principle. In addition we require {\it positivity} $\phi(a^+a)\geq 0$ and {\it faithfulness} in the sense that $\phi(a^{+}a)=0$ implies $a=0$. In addition we can require {\it normalizability} i.e. $\phi(1)=1$ for the identity $a=1$ of ${\cal{A}}({\cal{D}})$. In algebraic quantum field theory a linear form satisfying these properties is called {\it a state}. In particular we will say that $\phi$ is a {\it vector state} when we have the standard quantum mechanical representation $\phi(a)=\langle\Psi|a|\Psi\rangle$ for some vector state $|\Psi\rangle$ in $H_{QFT}$.

Before going on let us pause to think the causal structure of our background space-time. In the case of de Sitter and if we use the corresponding Penrose diagram we can identify the causal domain ${\cal{D}}$ of {\it one} observer with one of the two static patches. This leaves outside ${\cal{D}}$ the rest of the Penrose diagram that contains the {\it mirror} static patch as well as the two upper and lower triangular regions. These regions are causally very different from each other. While the two static patches are causally disconnected the observer can in principle create a fluctuation in her causal domain and this fluctuation can evolve in {\it global dS time} and to enter into the upper triangular region i.e. can exit the observer cosmological horizon. This phenomena simply reflects that translations in the {\it global dS time} are not keeping invariant the observer's causal domain i.e. the static patch. In other words, what we find is that while {\it global dS time} translations are {\it not defining  automorphisms of the observer algebra} their action can, in principle, be defined on {\it the full Hilbert space $H_{QFT}$}. The same can be said for the case of the eternal black hole space time where we can imagine fluctuations created outside that fall, evolving in global Schwarzschild time, into the black hole interior. 

\vspace{0.3cm}

After these preliminaries we could try to get for the {\it physics data} ${\cal{A}}({\cal{D}})$ and $H_{QFT}$ some information about the elements in $B(H_{QFT})$ that are not in ${\cal{A}}({\cal{D}})$. It can be shown that 
if ${\cal{A}}({\cal{D}})$ is a von Neumann factor \footnote{ Which means that satisfies i) the bicommutant relation i.e. ${\cal{A}}({\cal{D}})= {\cal{A}}({\cal{D}})^{''}$ for ${\cal{A}}({\cal{D}})^{'}$ the elements in $B(H_{QFT})$ commuting with all the elements in ${\cal{A}}({\cal{D}})$ and ii) has trivial center.}, then we get the representation of $B(H_{QFT})={\cal{A}}({\cal{D}})\otimes {\cal{A}}({\cal{D}})^{'}$. In other words {\it any element in $B(H_{QFT})$ which is not in ${\cal{A}}({\cal{D}})$ is in the commutant ${\cal{A}}({\cal{D}})^{'}$} \footnote{At this point the reader can wonder in what algebra is an observable localized outside the causal domain of the observer. What we know about this observable, let us say ${\cal{O}}$, is that for any state $|\psi\rangle$ in 
$H_{QFT}$ we cannot {\it distinguish}, using observables in ${\cal{A}}({\cal{D}})$, between $|\psi\rangle$ and ${\cal{O}}|\psi\rangle$ which implies, if ${\cal{O}}$ is unitary, that it belongs to the commutant ${\cal{A}}({\cal{D}})^{'}$. However, if we implement global time evolution in $H_{QFT}$ it could happens that the time evolved state ${\cal{O}}|\psi\rangle(t)$ can be distinguished, after some global time, from the state $|\psi\rangle(t)$ using measurements in ${\cal{A}}({\cal{D}})$.}.

In order to understand properly the physics meaning of the former statement it is convenient to recall the GNS construction. Given a state $\phi$ we can define a formal and {\it state dependent} Hilbert space $H^{\phi}$ that for $\phi$ faithful you can think as identical to the algebra ${\cal{A}}({\cal{D}})$ where to each element $a\in {\cal{A}}({\cal{D}})$ you associate a state $|a\rangle$ with {\it scalar product} $\langle a|b\rangle=\phi(a^{+}b)$. This is the GNS Hilbert space with a selected state $|1\rangle_{\phi}$ associated to the identity that is cyclic i.e. $H^{\phi}$ equal to the clousure of ${\cal{A}}({\cal{D}})|1\rangle$. The commutant of ${\cal{A}}({\cal{D}})$ defined {\it relative to $H^{\phi}$} has also the state $|1\rangle_{\phi}$ as cyclic and consequently $H^{\phi}$ is also equal to the clousure of ${\cal{A}}^{'}({\cal{D}})|1\rangle_{\phi}$ with ${\cal{A}}^{'}({\cal{D}})$ defined relative to $H^{\phi}$\footnote{This means that the state $|1\rangle_{\phi}$ is cyclic and separating.}.  In physics terms we can think of $H^{\phi}$ as a sort of "Fock space" generated by the observables in ${\cal{A}}({\cal{D}})$ or in the commutant. 

The key point of the GNS construction is that generically the Hilbert space $H^{\phi}$ will be {\it smaller} ( normally much smaller) than  $H_{QFT}$ and therefore the commutant {\it relative to $H_{QFT}$ will be much larger than the corresponding GNS commutant}. This implies that the notion of what observables account for local properties not accesible to the observer depends on what is the full Hilbert space $H_{QFT}$ \footnote{ As we will discuss in the next chapter in order to be sure that the GNS representation of a given von Neumann algebra ${\cal{A}}$ acting on $H$  defines itself a von Neumann algebra acting on $H^{\phi}$ we need to require that the state $\phi$ is {\it normal}. This in particular means that the state $\phi$ is a vector state in the augmented Hilbert space $H\otimes l^2(N)$.}

\vspace{0.3cm}

{\bf Modular time}. One crucial output of the GNS construction is to deliver a natural and deep connection between the tracial properties of the state $\phi$ and the {\it modular time}. This relation is known as Tomita Takesaki theory. For any positive and faithful state $\phi$ we can use the GNS construction to lift the adjoint action defined on ${\cal{A}}({\cal{D}})$ to $H^{\phi}$ as the operator $S$ \footnote{In general unbounded.} transforming $a|1\rangle_{\phi}$ into $a^{+}|1\rangle_{\phi}$. This action $S$ can be decomposed using the ({\it polar decomposition}) into a unitary transformation and a {\it "dilatation"}. For $\phi$ satisfying the trace property i.e. $\phi(ab)=\phi(ba)$ the {\it "dilatation"} component is {\it trivial} while for a {\it non tracial} state $\phi$ the dilatation is non trivial and defines {\it the modular time translation}. More specifically $S=J\Delta^{1/2}$ with $J$ unitary and with the {\it modular time} transformation generated by $\Delta_{\phi}^{it}$. This is the deep relation between traces and time.

At this point the reader can wonder why {\it trace states} lead to trivial modular time translations. The heuristic answer is easy to figure out. In standard quantum mechanics we represent a state $\phi(a)$ as $\langle \Psi|a|\Psi\rangle$ for some vector state $|\Psi\rangle$. This can be also represented as $tr(\rho_{|\Psi\rangle}a)$ with $\rho_{\Psi}=|\Psi\rangle\langle|\Psi|$. On the basis of this intuition we can identify states $\phi$ satisfying the trace property as those representing {\it a density matrix equal one}. Using now our statistical mechanics intuition we can think the modular Hamiltonian as $\ln\rho$ which for tracial states with $\rho=1$ leads to trivial modular Hamiltonian. However in case you consider {\it not tracial states} you can think in some non trivial associated $\rho$ and consequently in some non trivial modular dynamics.

\vspace{0.3cm}

{\bf Global integrable time and Connes' state.} As already stressed global time translations can be represented on $H_{QFT}$ although they {\it don't define automorphisms of} ${\cal{A}}({\cal{D}})$ and consequently they don't act on the GNS Hilbert space $H^{\phi}$. Hence, the most economical way, given $H^{\phi}$, to identify a bigger Hilbert space $H_{QFT}$ where {\it global time transformations} are acting, is to use as $H_{QFT}$ the space
\begin{equation}\label{Hilberts}
L^2({\mathbb{R}},H^{\phi})
\end{equation}
i.e. the space of square integrable functions $t\rightarrow |\Psi\rangle(t)$ from ${\mathbb{R}}$ into $H^{\phi}$ with the coordinate $t\in{\mathbb{R}}$ representing time. This space is special for several reasons. First of all we are introducing a sort of {\it integrable time} where by that we mean that the function $|\Psi\rangle(t)$ is square summable i.e.
\begin{equation}
\int_{-\infty}^{+\infty} |||\Psi\rangle (t)||^2
\end{equation}
is finite. This implies that the {\it constant state} is not in the Hilbert space (\ref{Hilberts}) and consequently that there is not a unitary transformation defining $|\Psi\rangle(t)$. Moreover we could think of the integrable time as a {\it complexity time} \footnote{We will not discuss in this note neither the meaning of $c(t)=:|||\Psi\rangle (t)||^2$ nor its integrated value $\int c(t)$. However it is clear that we can associate to any element in the Hilbert space (\ref{Hilberts}) and for any value of $t$ the corresponding {\it velocity} $v(t)=\frac{d|\Psi\rangle(t)}{dt}$ as well as a quantum length ( or {\it complexity}) relative to the natural Fubini Study metric on $H^{\phi}$ \cite{Gomezcomplex}.}. Once we define the Hilbert space (\ref{Hilberts}) the algebra acting on it is
\begin{equation}\label{algebra}
{\cal{A}}({\cal{D}})\otimes B(L^2(\mathbb{R}))
\end{equation}
In that way we are identifying the {\it global time} with {\it an extra quantum degree of freedom} with Hilbert space $L^2({\mathbb{R}})$.

This extra degree of freedom is characterized by a quantum {\it position} operator $t$ and its conjugated {\it momentum} that we will denote $H$ defining translations of the time {\it position} $t$. More precisely we can define $U(t)$ as the operator defining time translations on $L^2(\mathbb{R})$ with generator the {\it momentum} $H$. We will define Connes's state ( weight ) on the algebra (\ref{algebra}) by the linear form \cite{Connestaka}
\begin{equation}\label{connes}
\omega=\phi\otimes \bar \omega
\end{equation}
with $\bar\omega(x)= tr(x e^H)$ for $x\in B(L^2(\mathbb{R}))$. The important lesson of this construction is that in order to define integrable {\it time paths} on the GNS Hilbert space $H^{\phi}$ we need {\it to add an extra quantum degree of freedom}.

\vspace{0.3cm}

{\bf General Covariance and traces.} The action of time translations on the Hilbert space defined by (\ref{Hilberts}) is generated by the modular transformation associated to $\phi$ {\it times} the action $U(t)$ on $L^2(\mathbb{R})$. {\it General covariance} requires to identify as the algebra representing {\it physical observables} the elements invariant under the so defined time reparametrizations. This defines the centralizer ${\cal{C}}(\omega)$. This centralizer for ${\cal{A}}({\cal{D}})$ a type $III_1$ factor is a {\it crossed product} type $II_{\infty}$ factor. The great bonus of the construction is that on this centralizer i.e. on the algebra of {\it gauge invariant} observables, the weight $\omega$ (\ref{connes}) defines a trace $Tr$ i.e.
\begin{equation}
\omega(\hat a)=Tr(\hat a)
\end{equation}
for any $\hat a \in {\cal{C}}(\omega)$. This trace $Tr$ is neither finite nor normalizable. 

It could be worth to pause a moment to stress the logic underlying the previous construction. The crucial point is to extend the GNS Hilbert space $H^{\phi}$ about which you can think as the {\it minimal} representation of ${\cal{A}}({\cal{D}})$ into a model of $H_{QFT}$ hosting the action of {\it global time translations}. The {\it minimal} model of this Hilbert space is $L^2({\mathbb{R}})\otimes H^{\phi}$ that includes an {\it extra quantum degree of freedom} with Hilbert space $L^2({\mathbb{R}})$. This extra degree of freedom admits different {\it interpretations} as representing i) {\it the external observer} \cite{Witten2} ii) an added quantum {\it reference frame} \cite{Gomez2}\cite{Susskind2} for time in the spirit of \cite{reference} originally suggested in \cite{AS} iii) as an {\it external physical clock} \cite{Gomez3} or  iv) as an extra degree of freedom associated with local transformations defined on the boundary of the causal domain ${\cal{D}}$ \cite{Witten3}\cite{recent1,recent2,recent3}.

Our final physics goal is to implement on ${\cal{A}}({\cal{D}}) \otimes B(L^2({\mathbb{R}}))$ the requirement of {\it general covariance} and to define a {\it tracial linear form} on the so defined {\it gauge invariant} subalgebra. It is this trace the one used in \cite{Witten2,Witten3,Witten4} to define a QFT and weak gravity limit  version of Bekenstein Hawking entropy o Gibbons Hawking de Sitter entropy in the sense of (\ref{main}). Connes-Takesaki construction suggest to use as the natural candidate for this tracial state the one defined in (\ref{connes}). Here the reader should notice the strong {\it thermodynamic flavor} of the form $\bar\omega$ used to account for the contribution of the extra degree of freedom. Indeed $\bar\omega$ is the analog of the canonical way to define {\it thermal expectation values} relative to the $H$ generating the time translations $U(t)$ of {\it the added extra degree of freedom}.

\vspace{0.3cm}

{\bf Flow of weights and scale transformations.} The extra added degree of freedom allows us to define not only the $U(t)$ but also the canonically conjugated operator $V_s$ that generates "dual" transformations of "momentum" generated by the "position" operator. The action of $V_s$ on the tracial form $\omega$, known as {\it the flow of weigths},  is for ${\cal{A}}({\cal{D}})$ a type $III_1$ factor specially simple, namely just the scale transformation
\begin{equation}
(1\otimes V_s)\omega = \lambda \omega
\end{equation}
with $\lambda=e^{s}$. This simply corresponds to shift the extra " momentum" degree of freedom $H$ by a constant i.e. $H\rightarrow H+s$. This {\it scale transformation} acts as an automorphism on  the centralizer representing the {\it gauge invariant} ( with respect to general covariance time reparametrizations ) physical observable. Physically we can {\it break} this {\it scale symmetry} if we use a finite projection $p$ in the centralizer and we work with the projected type $II_1$ factor $p{\cal{C}}p$. Thus after projecting we {\it break the scale flow of weights} symmetry. What is the meaning of this breaking? This leads us to our final issue, namely the interpretation of the extra degree of freedom as a {\it physical clock}.

\vspace{0.3cm}

{\bf Clocks}. In order to intepret the extra degree of freedom as a physical clock we need to define, in addition to the {\it coordinate} and {\it momentum} ( $t$ and $H$ ) of the extra degree of freedom, an {\it hermitian and positive} Hamiltonian $H^{clock}$ formally defining $\dot t$ and $\dot H$ by the standard Heisenberg relation. For this clock we can formally define an {\it hermitian} operator $t^{clock}$ satisfying $[t^{clock},H^{clock}]=i\hbar$ but only on a selected subspace of clock states in $L^2({\mathbb{R}})$ characterized by $\dot t$ non vanishing and with small variance of $\dot t$ \footnote{Note that $t$ is the "position" operator of the extra degree of freedom while the meaning of the dot is defined by the Heisenberg equations for the hermitian hamiltonian $H^{clock}$.}.

A key proposal of this note, following previous attempts \cite{Gomez2,Gomez3}, is to define the {\it centralizer} ( representing the set of gauge invariant observables ) using a {\it clock} tracial form $\omega_{clock}$ where we replace the role of $H$ in the definition of $\bar \omega$ by the clock Hamiltonian $H^{clock}$ and we define the $tr$ entering into the definition of $\bar\omega$ by a suitable subspace of good clock states in $L^2({\mathbb{R}})$. The physics meaning of this proposal is in essence quite transparent. Namely we replace the "coordinate" $t$ of the extra degree of freedom into the {\it clock time} essentially defined by $\frac{t}{\langle \dot t \rangle}$ with $\dot t$ determined by $H^{clock}$. We expect that the so defined centralizer is the {\it physical} type $II_1$ factor both in the case of ${\cal{D}}$ the dS static patch as well as in the case where ${\cal{D}}$ is the causal domain of the eternal black hole asymptotic external observer. 

\vspace{0.9cm}

While adding an extra degree of freedom allows us to define a type $II_{\infty}$ centralizer for both the dS and the BH cases, once we promote this degree of freedom into a {\it physical clock} by adding a {\it positive and hermitian} clock Hamiltonian $H^{clock}$, we can define, again in both cases, a type $II_1$ factor. As said we could define this transit from type $II_{\infty}$ into type $II_{1}$ simply using a finite projection $p$ and defining the type $II_1$ factor as $p{\cal{C}}p$ for ${\cal{C}}$ the type $II_{\infty}$ centralizer.

Normally ( see \cite{Witten2,Witten3}) we distinguish between dS and the BH and we associate to the dS background the type $II_1$ factor defined by means of the finite projection $p$ while we assign to the BH the type $II_{\infty}$ factor. The physics motivation for that reflects the different nature, in both cases, of the added extra degree of freedom.

In particular the type $II_1$ nature of dS intends to explain why {\it empty dS} is associated with a {\it maximal entropy state}. This maximality accounts for the fact that any localized energy added in the static patch effectively reduces the GH entropy, making empty dS the maximal entropy state \cite{Torroba1,Torroba2,Torroba3,Torroba4}. This physics argument implies that for dS we must include the projection, or equivalently, to impose that the "momentum" $H$ of the extra degree of freedom must have a {\it positive spectrum}. Moreover the fact that the so defined type $II_1$ factor can be always interpreted as a subalgebra of the type $II_{\infty}$ factor effectively shows that in the dS case some elements of the type $II_{\infty}$ centralizer are {\it gauged away} as not representing physical observables i.e. physical fluctuations. From the point of view of the {\it flow of weights} we also {\it gauged away}, in the type $II_1$ case, the effect of this transformation.

 By contrast in the BH case we keep ourselves in the type $II_{\infty}$ factor on the intuitive basis that the BH entropy can be changed arbitrarily by adding extra matter with this physical phenomena reflecting the action of the flow of weights automorphism. However we could ask what is the meaning of the type $II_{\infty}$ description of dS as well as what is the meaning of a potential type $II_1$ description of the BH defined on a projected centralizer. To think in terms of {\it clocks} can shed some light on these questions.
 
 Indeed, we can, in principle, to promote, in both cases, the extra degree of freedom into a physical clock and in both cases we can use this physical clock, that by definition represents an extra degree of freedom with hermitian and {\it positive} Hamiltonian, to define the corresponding type $II_1$ factor.
 
The crucial point lies in distinguishing between an extra degree of freedom characterized by a position $t$ and momentum $H$ and with vanishing Hamiltonian $H^{clock}(t,H)=0$, that is the type $II_{\infty}$ case with $\bar \omega$ defined by $H$, from the type $II_1$ case where we add a non vanishing {\it positive} and hermitian Hamiltonian $H^{clock}(t,H)$ -- defining a physical time evolution i.e. $\dot t$ and $\dot H$ through the standard Heisenberg equations -- and where $\bar \omega$ is defined in terms of $H^{clock}$.

In words, while in the type $II_{\infty}$ case the Hilbert space is $L^2(\mathbb{R},H^{\phi})$
with coordinate time parametrizing ${\mathbb{R}}$ we now move into a different space that intends to represent $pL^2(\mathbb{R},H^{\phi})$ for some finite projection $p$ where now ${\mathbb{R}}$ represents the physical clock time and where translations in this time are generated by the hermitian and positive $H^{clock}$. 

Physically the difference between type $II_{\infty}$ and type $II_{1}$ can be understood thinking in the type $II_{\infty}$ extra degree of freedom as a formal {\it time observable} $t$ coordinate, with {\it conjugated momentum} $H$, of a trivial clock with vanishing {\it positive} hermitian Hamiltonian  and the type $II_1$ case as representing an extra degree of freedom with a non vanishing and positive hermitian Hamiltonian, $H^{clock}(t,H)$. 

In the case of dS we can visualize the type $II_{\infty}$ description thinking in the {\it extra degree of freedom} coordinate as the {\it inflaton vev $\phi_{inf}$ } with vanishing clock Hamiltonian i.e. with $\dot \phi_{inf}=0$. In Cosmology we will describe this situation in terms of a {\it flat} potential for the inflaton. For the BH case the type $II_{\infty}$ description uses the ADM mass $M_{ADM}$ as natural coordinate with vanishing Hamiltonian i.e. $\dot M_{ADM}=0$. In both case the {\it flow of weights} automorphism of the type $II_{\infty}$ factor generates shifts of either the inflaton $\phi_{inf}$ or the ADM mass $M_{ADM}$.

The type $II_{1}$ version, in both cases, requires to add the clock Hamiltonian inducing $\dot \phi_{inf}$ for the dS case or $\dot M_{ADM}$ for the BH case. Physically this type $II_1$ version is different in both cases. In the dS case it leads to a type $II_1$ by adding a non vanishing $\dot \phi_{inf}$ that is the standard slow roll inflationary parameter and {\it not} adding {\it by hand} a projection \footnote{Note that imposing positivity for the momentum of the extra degree of freedom is not as innocent as it could appear at first sight if the coordinate extra degree of freedom has some concrete physical meaning. To use in the case of dS as extra degree of freedom coordinate the vev $\phi_{inf}$ seems to us the most natural guess. Obviously in this case the constrain projecting on positive momentum is far from trivial.} In the BH case the situation is potentially more interesting because leads to a possible type $II_1$ description of BH's once we consider a non vanishing $\dot M_{ADM}$ that is related to {\it Hawking temperature}. To distinguish between the type $II_1$ factors obtained by including a {\it clock} degree of freedom we need to work out the way these factors are acting on different Hilbert spaces. The simplest way to distinguish these different type $II_1$ implementations can be done in terms of the {\t Murray-von Neumann coupling constant}. In the last section we will briefly discuss the physics meaning of this coupling constant for a type $II_1$ description of black holes and we will define a natural associated Page's curve \cite{Page} \footnote{A formal relation between black holes and de Sitter can be defined if the type $II_1$ factor associated to de Sitter is acting on a Hilbert space $H$ with Murray-von Neumann coupling constant equal to infinity. In such a case we can formally identify the corresponding black hole type $II_{\infty}$ factor with {\it the commutant}, relative to $H$, of the type $II_1$ de Sitter factor}.

\section{Traces: some basic mathematical properties}
In this section we will review most of the mathematical tools used in the rest of the paper. The reader can find most of the mathematical results on von Neumann algebras used in this note in the  review by V.Jones \cite{Jones} and in \cite{wittenmedal}.
\subsection{Some preliminary comments on projections}
Let $H$ be a separable Hilbert space with $\dim H=\infty$. We will think $H$ as the Hilbert space of a generic quantum mechanical system. Let $B(H)$ be the algebra of bounded operators acting on $H$ \footnote{Recall that an operator $a$ is in $B(H)$ if for any $|\psi \rangle \in H$ there exist a number $k$ such that $||a|\psi \rangle ||\leq k |||\psi\rangle||$ for any $|\psi\rangle\in H$. The smaller value of $k$ is the norm of $a$.}. A von Neumann algebra ${\cal{A}}$ is a self adjoint subalgebra of $B(H)$ 
i.e.  ${\cal{A}}^{+} = {\cal{A}}$ for $a^{+}$ the adjoint, {\it with unit } and such that ${\cal{A}}^{''}={\cal{A}}$ for ${\cal{A}}^{'}$ the commutant of ${\cal{A}}$ in $H$ i.e. the set of elements in $B(H)$ commuting with all elements in ${\cal{A}}$. A {\it factor} is a von Neumann algebra with trivial center i.e. elements of the center are multiples of the identity. Note that $B(H)$ itself is a factor. Moreover for $H=H_1\otimes H_2$ we easily see that $B(H_1)\otimes 1$ is a factor with $1\otimes B(H_2)$ as its commutant. 

We will say that a factor ${\cal{A}}$ acting on $H$ is {\it type $I$} if there exist a decomposition of $H=H_1\otimes H_2$ such that ${\cal{A}}=  B(H_1)\otimes 1$ with commutant ${\cal{A}}^{'}= 1\otimes B(H_2)$. For any factor ${\cal{A}}$ acting on $H$ the algebra $B(H)$ {\it splits} as $B(H)={\cal{A}}\otimes {\cal{A}}^{'}$ but {\it only} for a type $I$ factor this split property extends to the Hilbert space itself $H=H_1\otimes H_2$ with ${\cal{A}}= B(H_1)\otimes 1$ and ${\cal{A}}^{'}= 1\otimes B(H_2)$. To understand the logic underlying these definitions is good to recall  that for  $\dim H$ {\it finite}  any self adjoint subalgebra $A$ with unit satisfies the bicommutant relation $A^{''}=A$.

Physically for a given quantum system ${\cal{S}}$ with Hilbert space $H$ we can think of any factor ${\cal{A}}$ as defining an algebraic {\it partition} of $B(H)$ into two {\it subsystem commuting algebras}, namely ${\cal{A}}$ and ${\cal{A}}^{'}$. That ${\cal{A}}$ is a type $I$ factor implies, among other things, the existence in $H$ of pure states with {\it zero entanglement} between both {\it algebraically defined} subsystems.

{\it Projections} are operators in $B(H)$ satisfying $p^2=p$. In quantum mechanics we normally interpret projections as representing Yes/No questions that can be answered performing a certain complex measurement. The projection represents the corresponding observable with eigenvalues $0,1$. Projections can be ordered defining $p\leq q$ if {\it algebraically} $pH \subseteq qH$. {\it Minimal projections} are projections $p$ for which if $q\leq p$ then $q$ is either zero or $p$. In quantum mechanics we associate {\it pure states} to minimal projections i.e those for which $pH$ is one dimensional.

{\it Isometries} are elements $u\in B(H)$ such that $u^{+}u=1$ being unitary if $u^{+}u=uu^{+}=1$. {\it Partial isometries} are those for which $u^{+}u$ is a projection. If $\dim H=\infty$ we can define a non trivial {\it equivalence relation} between projections by $p \approx q$ if there exist a partial isometry $u$ such that $uu^{+}=p$ and $u^{+}u=q$. Moreover we can define a partial order $p \precsim q$ if there exist a partial isometry $u$ such that $p=uu^{+}$ and $u^{+}u \leq q$. If $\dim H=\infty$ we have {\it infinite projections} as those for which $pH$ is itself infinite dimensional although it could be a subspace of $H$. We characterize these infinite projections as those projections $p$ for which exist another projection $q\neq p$ such that $p \approx q$ for some $q\leq p$. Note that $\approx$ among projections extends the notion of equivalent projections to infinite dimension. Moreover note that for $p \approx q$ there exist a partial isometry $u$ such that $p=u^{+}qu$. 

Now for a von Neumann algebra ${\cal{A}}$ acting on $H$ and any projection $p\in{\cal{A}}$ it can be proved \cite{Jones} that $p{\cal{A}}p$ is a von Neumann algebra acting on $pH$ and that the commutant $(p{\cal{A}}p)^{'}$ on $pH$ is ${\cal{A}}^{'} p$. 

Finally we can define a {\it complete set of mutually orthogonal projections} $p_i$ with $i=1,....$ by the condition $p_ip_j=\delta_{i,j}p_i$ and
\begin{equation}
\sum_ip_{i} = 1
\end{equation}
For ${\cal{A}}$ a von Neumann factor
 acting on $H$ and $p_i$ a complete set of mutually orthogonal projections in ${\cal{A}}$ the factor ${\cal{A}}$ can be represented as
 \begin{equation}\label{formula}
 p_1{\cal{A}}p_1\otimes B(l^2(N))
 \end{equation}
 for $l^2(N)$ the Hilbert space of square summable sequences. 

\subsection{Traces}
Given a separable Hilbert space we can define a basis $\xi_i$ and the trace of any operator $a\in B(H)$ as $\sum_i \langle \xi_i a \xi_i\rangle$. This quantity is not well defined for all elements in $B(H)$, for instance the so defined trace of the identity is infinity. For $B(H)$ we can define the set of {\it trace class} operators as those with $tr(|a|)$ finite. Let us denote this set $B(H)_{1}$. For $dim H=\infty$ it can be proved that $B(H)_{1}$ is not an algebra but the so called {\it predual} of $B(H)$ i.e. a Banach space such that the dual $B(H)_{1}^{*}$ is $B(H)$. Obviously this set of trace class operators cannot be an algebra since the identity is not part of $B(H)_{1}$. It is easy to check that $B(H)_{1}$ is a {\it two sided ideal} in the sense that for any $a\in B(H)_{1}$ and any $b\in B(H)$ we have that $ab$ as well as $ba$ are in $B(H)_{1}$. One of the key physical problems we will face is to find a {\it sub algebra} ${\cal{A}}$ of $B(H)$ that is a von Neumann algebra acting on $H$ and that it is equipped with a finite trace. Obviously $B(H)_{1}$ is not an answer since this space is not an algebra.

 We will define a {\it trace state} on a von Neumann factor ${\cal{A}}$ as a linear form $tr:{\cal{A}}\rightarrow {\mathbb{C}}$ satisfying i) $tr(ab)=tr(ba)$ ii) positivity $tr(a^{+}a)\geq 0$, iii) faithfulness i.e. $tr(a^{+}a)=0$ implies $a=0$ and iv) normalizability i.e. $tr(1)=1$. Generically we will say that a linear form $\phi$ on ${\cal{A}}$ is {\it a state} if it satisfies the conditions of positivity faithfulnes and normalizability but {\it not} the trace condition i) \footnote{Note that positivity implies that $\phi(a^+)=\bar{\phi(a)}$.}. 
 
 Now we will look for von Neumann factors in $B(H)$ for which it exists a trace satisfying the former conditions. A priori is far from clear the existence, for $\dim H=\infty$, of any factor having a trace. Intuitively since the factor must contain algebraically the identity we need to define the linear form $tr$  in such a way that the trace of the {\it algebraic identity} $1$ of ${\cal{A}}$ satisfies  $tr(1)=1$ for $\dim H=\infty$. In particular note that for the factor $B(H)$ {\it there is not any linear form} $tr$ satisfying the trace properties.

Before looking for such factors let us note that a trace in the former sense is naturally defined on the equivalence classes of projections. Namely $tr(p)=tr(q)$ is $p\approx q$ since $tr(p) = tr(u^{+}qu) = tr(uu^{+}q)=tr(q)$ where we used that $p=u^{+}u$ and $q=uu^{+}$. 

Factors having a trace in the former sense are known as type $II_1$ factors and they were originally discovered by Murray and von Neumann. As we will see these factors are specially interesting in physics since they capture a natural notion of {\it finiteness}. In particular for any projection $p$ in a type $II_1$ factor $tr(p)\in[0,tr(1)=1]$.  Recall that the physics question we are facing is to find a factor ${\cal{A}}$ acting on $H$ equipped with a finite trace and such that $B(H)={\cal{A}}\otimes {\cal{A}}^{'}$.

Probably the reader will be wondering why for a given decomposition of $H=H_1\otimes H_2$ we are not happy to describe the corresponding associated subsystems with Hilbert spaces $H_1$ and $H_2$ in terms of the trace class operators $B(H_1)_{1}$ and $B(H_2)_{1}$. The reason, as already stressed, is that these sets of trace class operators are not {\it algebras}. Specially interesting operators in $B(H)$ are the Hilbert Schmidt (HS) operators defined as those operators $a\in B(H)$ such that $a^{+}a$ is trace class. Again the space of HS operators don't define an algebra\footnote{In particular if for $H=H_1\otimes H_2$ we consider the type $I_{\infty}$ factor $B(H_1)\otimes 1$ with commutant $1\otimes B(H_2)$ there is no non zero linear form satisfying the trace property if dimension of $H$ is infinity. To use $B(H_1)_1$ or the space of HS operators can provides a trace at the price of spoiling the algebra structure.}. Note that for the TFD construction with $H_1$ and $H_2$ two identical copies there is no non zero linear form trace $tr(O)$ for operators $O$ in the type $I_{\infty}$ factor $B(H_1)\otimes 1$ if $dim H_1=\infty$.

\subsection{The q-bit model of type $II_1$ factors}
As said we are interested in defining a factor ${\cal{A}}$ acting on $H$ having a {\it trace state}. We will define a q-bit model of such a factor. 

Let us define the algebra $A =M_2(\mathbb{C})$ i.e. complex two by two matrices. Define the {\it infinite algebraic tensor product} $A_{\infty}$ as $\bigcup A_n$ with $A_n= \otimes^{n}A$ with the identity of $A_{\infty}$ as $1=1\otimes 1 \otimes 1 ..$. Using the trace in $A$ i.e. the natural trace in $M_2(\mathbb{C})$, we can define a pre-Hilbert space ${\cal{H}}_{tr}$  by the sesquilinear form
\begin{equation}\label{formm}
\langle a|b\rangle = tr(a^{+}b)
\end{equation}
for $a$ generated by $a_1\otimes a_2\otimes ..a_n\otimes 1\otimes 1....$ and similarly for $b$ where we have a {\it finite} number of matrices $a_i\in A$ different from $1$\footnote{Note that (\ref{formm}) becomes the finite product $\prod tr(a_i^{+}b_i)$ over the finite number of elements in $a$ and $b$ different from the identity}. The Hilbert space $H_{tr}$ can be defined as the topological completion of the so defined pre Hilbert space. The previous construction is a typical example of GNS construction where you associate to any element $a$ defined by the tensor product of a {\it finite} number of matrices in $M_2(\mathbb{C})$ different from the identity, a vector $|a\rangle$ and where you use the {\it finite product} of the traces, as given in (\ref{formm}), to define a pre Hilbert space. We will denote $|1\rangle$ the vector associated with the identity.

A basis of matrices in $M_2(\mathbb{C})$ are the two diagonal matrices $d_i$ with $i=1,2$ and with $d_1=1$ and the matrices $e_{i,j}$ with $i\neq j$ and the only non vanishing entry $(i,j)$ equal to one. Thus a basis of $H_{tr}$ is 
\begin{equation}
\otimes v_i |1\rangle
\end{equation}
with $|1\rangle$ the state associated with the identity $1=1\otimes 1\otimes ...$ and with $v_i=1$ for $i>K$ for $K$ arbitrarily large but {\it finite}. The action of ${\cal{A}}_{\infty}$ on $H_{tr}$ is defined by {\it left multiplication} $a|b\rangle = |ab\rangle$. The {\it commutant} ${\cal{A}}_{\infty}^{'}$ is defined by {\it right multiplication} i.e. $a'|b\rangle = |ba\rangle$ for $a'\in {\cal{A}}_{\infty}^{'}$. Note also that in this representation where  we are acting on $H_{tr}$ the algebras ${\cal{A}}_{\infty}$ and ${\cal{A}}_{\infty}^{'}$ are two copies of the same algebra. 

Physically we can think $M_2(\mathbb{C})$ as acting on the four dimensional Hilbert space representing two q-bits \cite{wittenmedal}. Thus $H_{tr}$ describes an infinite set of pairs of q-bits. The basis elements $d_i$ and $e_{i,j}$ can be interpreted as {\it elementary operations} on the pair of q-bits. Hence elements in $H_{tr}$ represent the different states we can obtain acting with a {\it finite number} of elementary operators on the {\it ground state} represented by $|1\rangle$.

Now we can {\it lift the  adjoint} i.e. the $+$ operation of $M_2(\mathbb{C})$ to $H_{tr}$ and to define the {\it modular operator} $S_{tr}$ as
\begin{equation}\label{modularr}
S_{tr} a |1\rangle =a^{+}|1\rangle
\end{equation}
We easily see that $S_{tr}e_{i,j} |1\rangle = e_{j,i}|1\rangle$ implying $S_{tr}^2=1$ and
\begin{equation}\label{modular}
S_{tr}{\cal{A}}_{\infty}S_{tr} = {\cal{A}}_{\infty}^{'}
\end{equation}
If we define the modular operator $\Delta_{tr}= S_{tr}S_{tr}^{+}$ with $S_{tr}^{+}$ defined by (\ref{modular}) but for $a\in {\cal{A}}_{\infty}^{'}$ we get
\begin{equation}
\Delta_{tr} = 1
\end{equation}
Thus we have defined an algebra ${\cal{A}}$ as the completion of ${\cal{A}}_{\infty}$ acting on $H_{tr}$ {\it equipped with a trace state}, namely
\begin{equation}\label{definition}
tr(a)=\langle 1|a|1\rangle
\end{equation}
satisfying all the required properties: positivity, faithfulness, normalizability and 
trace property. This is the q-bit model of a type $II_1$ factor. The main lesson for our future discussion is that the {\it modular dynamics defined by $\Delta_{tr}$ is trivial.}

Note that once we have defined the {\it trace state} on ${\cal{A}}$ by (\ref{definition}) we can associate with any vector state $|\Psi\rangle$ in $H_{tr}$ the corresponding density matrix $\rho_{\Psi}$ by
\begin{equation}
tr(\rho_{\Psi} a)=\langle \Psi|a|\Psi\rangle
\end{equation}
which implies that $\rho_{1}=1$ i.e. the density matrix associated with the GNS ground state $|1\rangle$ is just the identity. In physics terms this can be interpreted as saying that $|1\rangle$ is a {\it flat entanglement state}. Note that this {\it flat entanglement} is a consequence of the trivial modular dynamics i.e. $\Delta_{tr}=0$.

\subsection{The type $III$ relative}
An obvious question, after the previous description, is why not to use, instead of the sesquilinear form (\ref{formm}), a different one, namely
\begin{equation}\label{newform}
\langle a|b\rangle_{h} = tr(ha^{+}b)
\end{equation}
with $h$ diagonal, invertible and of trace equal $1$. Any matrix satisfying this property can be written as $h_{11}= \frac{1}{1+\lambda}$, $h_{22}= \frac{\lambda}{1+\lambda}$ and $h_{1,2}=h_{2,1}=0$ for some $\lambda \in [0,1]$. Obviously we can repeat for ${\cal{A}}_{\infty}$ defined above the whole construction but defining now $H_{tr}$ using (\ref{newform}). We can denote this space as $H_{\lambda}$ and the sesquilinear form (\ref{newform}) as 
$\langle a|b\rangle_{\lambda}$. The changes induced by this apparently minor modification are basically two:
\begin{itemize}
\item The linear form defined now by $\phi_{\lambda}(a)=:\langle 1|a|1\rangle_{\lambda}$ although faithful and positive is {\it not anymore satisfying the trace property}. This is obvious from (\ref{newform}) if $h$ is different from the identity.
\item The lift of the adjoint to $H_{\lambda}$, defined by some $S_{\lambda}$ satisfying the analog of (\ref{modularr}) leads, by {\it polar decomposition}, to a {\it non trivial} modular operator $\Delta_{\lambda}$. In the q-bit model the modular operator $\Delta_{\lambda}$ can be obtained as the tensor product of the $\Delta^{(2)}_{\lambda}$ defined for just one copy of $M_2(\mathbb{C})$ for the linear form defined by $\phi_{\lambda}(a)=tr(h_{\lambda}a)$ for $a\in M_2(\mathbb{C})$. It is easy to see that 
\begin{equation}
\Delta^{(2)}_{\lambda} (e_{1,2}) = \frac{1}{\lambda} e_{1,2}
\end{equation}
and 
\begin{equation}
\Delta^{(2)}_{\lambda} (e_{2,1}) = \lambda e_{2,1}
\end{equation}
\end{itemize}
The so defined factor is known as a $III_{\lambda}$ Powers factor. For this factor there is not trace state and {\it the modular dynamics is non trivial}.

\subsubsection{Type $III$ time as double scaling}
In order to gain some intuition let us see how the modular operator $\Delta_{\lambda}$ defined by the tensor product of the $\Delta^{(2)}_{\lambda}$ is acting on elements in $H_{\lambda}$. Recall that $H_{tr}$ and $H_{\lambda}$ are algebraically the same space. They differ as Hilbert spaces since they are equipped with different scalar products, namely one defined by (\ref{formm}) and the other by (\ref{newform}). Basis elements in $H_{\lambda}$ are obtained by acting on the ground state $|1\rangle_{\lambda}$ with a {\it finite number} of elements $e_{i,j}$ and $d_2$ i.e. are defined as
\begin{equation}\label{prehilbert}
\otimes a_n|1\rangle_{\lambda}
\end{equation}
with $a_i=1$ for $i>K$ for some large but finite $K$. Although the Hilbert space is defined after completion let us consider the preHilbert space (\ref{prehilbert}) with some large but undefined finite $K$. If we naively act with the modular operator $\Delta_{\lambda}$ on one of these states defined by $a_n=e_{2,1}$ for all $n\leq K$ we get
\begin{equation}
\Delta_{\lambda}(\otimes a_n|1\rangle_{\lambda})= \lambda^K (\otimes a_n|1\rangle_{\lambda})
\end{equation}
Thus if we define a {\it modular Hamiltonian} $h_{\lambda}$ as $\ln \Delta_{\lambda}$ the eigenvalue of $h_{\lambda}$ on this state will be $K\ln(\lambda)$ where a priori $K$ can be any finite and arbitrarily large integer number. Although this {\it physics approach} to the spectrum of the modular Hamiltonian is just heuristic let us try to proceed a bit further. 

One aspect of the previous construction that probably could sounds paradoxical is that for $\lambda=1$ there is not real difference between the construction of the type $II_1$ factor above and the construction of the type $III_{\lambda=1}$ factor. However this is not necessarily so. Indeed while for the type $II_1$ factor the modular hamiltonian is trivial and identically zero we could use a {\it double scaling limit} to define the limit $\lambda=1$ of the $III_{\lambda}$ factor. In particular we can {\it suggest} to define the modular spectrum by the double scaling limit
\begin{equation}\label{spec}
\lim_{K=\infty,\lambda=1} K\ln(\lambda)
\end{equation}
that we can take as an arbitrary positive real number. If we do that we get a non trivial modular Hamiltonian $h_{\lambda=1}$ with spectrum defined by (\ref{spec}) i.e. ${\mathbb{R}}^{+}$. Note that the so defined modular hamiltonian $h_{\lambda=1}$ is not part of the algebra. 

The key point of this suggestion is that in this double scaling limit we get, by contrast with the type $II_1$ case, a {\it non trivial modular dynamics} and consequently a non trivial {\it modular time evolution}. Morally speaking we observe that {\it time emergency} in this algebraic setup appears when we send $K$, representing the number of elementary operations we do on the ground state of an infinite set of pairs of q-bits, to infinity. In a certain heuristic way:

{\it In the type $III_1$ case time emerges when we push the dimension of the pre Hilbert space $H_{\lambda}$ i.e. the cardinality of the basis, to infinity and we define the limit $\lambda=1$ by a double scaling limit}.

The previous {\it q-bit} construction of the type $III_{\lambda}$ factors is known as {\it Powers construction}. Tomita Takesaki theory extends this construction to the GNS construction for a von Neumann algebra $M$ defined by a state $\phi$ {\it not satisfying the trace property}. In such a case the extension of the adjoint map in $M$ and $M^{'}$ to the GNS Hilbert space $H_{\phi}$ leads to unbounded operators $S_{\phi}$ and $S_{\phi}^{+}$ with polar decompositions $S_{\phi}=J_{\phi}\Delta_{\phi}^{1/2}$ with non trivial $\Delta_{\phi}=S_{\phi}^{+}S_{\phi}$.

\subsubsection{Murray-von Neumann dimension: the coupling constant}
Intuitively type $II_1$ flat entanglement in the sense of $\rho_1=1$ for the {\it cyclic} state playing the role of ground state is interpreted, since $tr(\rho_1)$ is normalizable for $tr$ the trace of the type $II_1$ factor, as implying a finite dimensional Hilbert space. This intuition is based on a microcanonical interpretation of $\rho_1$. When we move into the type $III_1$ factor we don't have even a trace to be used to define some formal dimension of the Hilbert space.

Nevertheless for a generic type $II_1$ factor ${\cal{A}}$ acting on an infinite dimensional Hilbert space $H$ we can give meaning to a notion of {\it dimension} of $H$ relative to ${\cal{A}}$ where we effectively measure the "size" of $H$ {\it relative} to the "size" of the GNS Hilbert space $H_{tr}$. This dimension is denoted $d_{{\cal{A}}} H$. To define this dimension we consider the Hilbert space ${\cal{H}} = H\otimes H_{tr}\otimes l^2(N)$ and we look for a projection $p$ such that $p{\cal{H}} = H$. Now for ${\cal{A}}$ a type $II_1$ factor there exist an isometry
\begin{equation}\label{isometry}
u: H\rightarrow H_{tr} \otimes l^2(N)
\end{equation}
with $p=u^{+}u$. In these conditions we define
\begin{equation}
d_{{\cal{A}}} H = tr(u^{+}u)
\end{equation}
For the case $H$ is the GNS Hilbert space $H_{tr}$ we get $d=1$. Note that this dimension can be arbitrarily large i.e. is in $[0,+\infty]$. For instance if $H=H_{tr}\otimes l^2(N)$ we get $d=\infty$. If we think of $d$ as some measure of the Hilbert space dimension we could define a formal {\it entropy} $\tilde S({\cal{A}},H) = log(d_{{\cal{A}}} H)$. 

An interesting property of $d_{{\cal{A}}} H$ is that it informs us about the existence of a cyclic or separating state for ${\cal{A}}$. In particular if $d_{{\cal{A}}} H\in [0,1]$ then there exist a cyclic state and if $d_{{\cal{A}}} H)\in [1,+\infty]$ there exist a separating state or equivalently a cyclic state for the commutant. Only for the special case $d_{{\cal{A}}} H=1$ corresponding to the GNS representation $H_{tr}$ we have a cyclic and separating state.

Moreover if $d_{{\cal{A}}} H=\infty$ the commutant  ${\cal{A}}^{'}$ is a type $II_{\infty}$ factor thus ${\cal{A}}$ is type $II_{\infty}$ if
\begin{equation}
d_{{\cal{A}}^{'}} H =\infty
\end{equation}
Hence if $d_{{\cal{A}}} H=\frac{1}{d_{{\cal{A}}^{'}} H}$ the limit case where ${\cal{A}}$ is a type $II_{\infty}$ factor corresponds to the formal limit $d_{{\cal{A}}} H=0$ \footnote{We will use these properties of $d$ in the last section to identify a Page's curve for a type $II_1$ version of BH's. Note also that in \cite{Witten2} the limit case $d_{{\cal{A}}} H=0$ is interpreted, in the weak gravity limit, as representing the case with no observer in the second patch.}.

\subsection{Type $II_{\infty}$ factors}
A type $II_{\infty}$ factor is defined as
\begin{equation}
{\cal{A}} \otimes B(H)
\end{equation}
for ${\cal{A}}$ a type $II_1$ factor and $H$ a Hilbert space of {\it infinite dimension}. It would be convenient to distinguish the Hilbert space on which the type $II_1$ factor is acting that we generically denote $H_{{\cal{A}}}$ and the Hilbert space $H$. For instance when we define the type $II_1$ factor as acting on the GNS Hilbert space $H_{tr}$ defined by the type $II_1$ trace described in the previous sections, then the type $II_{\infty}$ factor ${\cal{A}} \otimes B(H)$ is acting on the Hilbert space $H_{tr}\otimes H$.
\subsubsection{Trace on $II_{\infty}$}
Using the trace for the type $II_1$ factor ${\cal{A}}$ we can define a positive linear form $Tr$ on ${\cal{A}} \otimes B(H)$ satisfying the trace property. This can be done identifying the elements in ${\cal{A}} \otimes B(H)$ for a basis $\xi_i$ of $H$ as matrices valued in ${\cal{A}}$ i.e. matrices $(x_{i,j})$ with entries $x_{i,j} \in {\cal{A}}$. Once this is done we define
\begin{equation}\label{Trace1}
 Tr(a) = \sum_i tr(x_{ii})
\end{equation}
where $a\in {\cal{A}} \otimes B(H)$ and with "matrix" $x_{i,j}$ the corresponding representative of $a$ for a given basis of $H$. Contrary to the type $II_1$ trace this type $II_{\infty}$ trace is not finite $Tr(1)=\infty$ and cannot be normalized. 

Using (\ref{Trace1}) we can classify {\it projections} in ${\cal{A}} \otimes B(H)$ depending on the value of $Tr(p)$. We will say that $p$ is finite if $Tr(p)$ is finite \footnote{Note that now $Tr(p)\in [0,+\infty]$.}. Now comes an important result for our future physics discussion. For any infinite ${\cal{B}}$ factor if there exist a projection $p$ in ${\cal{B}}$ such that $p{\cal{B}}p$ is a type $II_1$ factor then ${\cal{B}}$ is a type $II_{\infty}$ factor such that
\begin{equation}
{\cal{B}} = p{\cal{B}}p \otimes B(H)
\end{equation}
for $H$ infinite dimensional.
 The ingredient needed to prove this statement is that the existence of $p$ satisfying that $p{\cal{B}}p$ is a type $II_1$ factor implies the existence of a complete family of mutually orthogonal equivalent projections $p_i$ in ${\cal{B}}$ satisfying $p_ip_j=\delta_{i,j}p_i$, $p_i\approx p$ and $\sum_ip_i=1$. From here it follows ( see (\ref{formula}))
that ${\cal{B}}$ is isomorphic to
\begin{equation}
p{\cal{B}}p \otimes B(H)
\end{equation}
and consequently is a type $II_{\infty}$ factor. 

In other words for any type $II_{\infty}$ factor ${\cal{B}}$ we can represent this factor as ${\cal{B}}= {\cal{A}} \otimes B(H)$ for some type $II_1$ factor ${\cal{A}}$ and to define the associated $Tr$. Reciprocally for any finite projection $p\in {\cal{A}}\otimes B(H)$ i.e. with $Tr(p)$ finite, we can define a type $II_1$ factor as $p({\cal{A}} \otimes B(H))p$.

At this point a very natural question is how the type $II_1$ factor defined as $p{\cal{B}}p$ for $p\in {\cal{B}}$ depends on the used projection $p$. The way to address this problem is to compare the $Tr$ on ${\cal{B}}$ defined by (\ref{Trace1}) -- using the decomposition ${\cal{B}}=p{\cal{B}}p\otimes B(H)$ -- for different projections $p\in{\cal{B}}$ for which $p{\cal{B}}p$ is a type $II_1$ factor. Obviously if the trace $Tr$ on ${\cal{B}}$ would be unique we could expect that the different type $II_1$ factors $p{\cal{B}}p$ will not depend on $p$. However this is not the case. Indeed the trace $Tr$ defined by the decomposition of ${\cal{B}}$ into the product of a type $II_1$ factor and $B(H)$ is only unique {\it up to rescaling} by some $\alpha >0$. This fact reflects the existence of an automorphism $\alpha$ of ${\cal{B}}$ such that

\begin{equation}\label{automor}
Tr(\alpha(a)) = \alpha Tr(a)
\end{equation}
for $\alpha >0$. Hence we could think of the existence of a one parameter family of associated type $II_1(\alpha)$ factors such that the traces $Tr(\alpha)$ defined by the representation ${\cal{B}}=II_1(\alpha)\otimes B(H)$ are related by re scalings. This discussion will have some interesting physical meaning whose discussion we postpone 
to the next section.

\subsection{Traces and centralizers}
In previous sections we have defined $tr$ for a type $II_1$ factor and $Tr$ for a type $II_{\infty}$ factor but for a type $III$ factor $M$ acting on a Hilbert space $H$ we don't have any faithful and normal state $\phi$ satisfying the trace property \footnote{A {\it normal state} is a linear form $\phi$ positive, faithful and satisfying complete additivity. By that we mean that for a family of mutually orthogonal projections $p_i$ we have $\phi(\vee p_i) = \sum_i \phi(p_i)$. Normality is important in order to be sure that the GNS representation defined by $\phi$ is itself a von Neumann algebra. Moreover for a factor $M$ acting on a Hilbert space $H$ a normal state becomes a {\it vector state} in an extended Hilbert space $H\otimes l^2(N)$.}. Thus, a natural question is: can we find a subalgebra of $M$ such that $\phi$ reduced to this subalgebra is a trace? This is equivalent to ask if inside $M$ we can find a subalgebra that is a type $II_1$ or a type $II_{\infty}$ factor. 

The answer to this question is related with the so called KMS condition. For a given normal and faithful state $\phi$ defined on $M$ we can define: i) the corresponding GNS representation on a Hilbert space $H_{\phi}$ ii) the modular operator $\Delta_{\phi}$, by lifting the adjoint action of $M$ to $H_{\phi}$ and iii) the corresponding  modular automorphism $\sigma_{\phi}^{t}(a) =\Delta_{\phi}^{it} a \Delta_{\phi}^{-it}$ defined by $\Delta_{\phi}$. A crucial result is that \footnote{Assuming $\sigma_{\phi}^{t}$ can be extended to imaginary times.}
\begin{equation}
\phi(ab)=\phi(b\sigma_{\phi}^{i}a)
\end{equation}
Thus it follows that $\phi$ defines a {\it trace} on the set of elements $a\in M$ invariant under the action of $\sigma_{\phi}(t)$. This set  defines  {\it the centralizer} ${\cal{C}}(\phi,M)$.

Denoting $h_{\phi}$ the modular Hamiltonian defining $\Delta_{\phi}$ the centralizer is simply the invariant subalgebra $M^{h_{\phi}}$ i.e. those elements in $M$ commuting with $h_{\phi}$. For $M$ a type $III_{1}$ factor this invariant subalgebra is trivial \footnote{Due to the ergodicity of the modular action.} and reduces to the multiples of the identity. So, how to get a non trivial centralizer ?

\subsubsection{Connes-Takesaki theorem}

Recall that our task is to find for a factor $M$ a state ( actually a weight ) $\omega$ such that the {\it centralizer} of $\omega$ is either a type $II_{\infty}$ or a type $II_1$ factor. In order to discover the desired factor $M$ let us recall Connes-Takesaki theorem \cite{Connestaka}. 

Let us consider a von Neumann factor ${\cal{M}}$ and a weight $\phi$ i.e. a normal state $\phi$ from the positive elements in ${\cal{M}}$ into $[0,+\infty]$. Using $\phi$ let us define the corresponding modular automorphism $\sigma_{\phi}^{t}$ with modular generator $h_{\phi}$. 

Let us now consider the type $I_{\infty}$ factor defined by $B(L^2(\mathbb{R}))$ i.e. the bounded operators on $L^2(\mathbb{R})$. Using $t$ to parametrize the real line $\mathbb{R}$, elements in $L^2(\mathbb{R})$ are  just square integrable functions $f(t)$. Let us define the unitary representation on $L^2(\mathbb{R})$ of {\it translations} on $\mathbb{R}$ i.e.
\begin{equation}
U_{t_0} f(t) = f(t+t_0)
\end{equation}
for reasons that will be clarified later let us denote $H^r$ the generator of $U_t$. On $L^2(\mathbb{R})$ we can define the conjugated unitary transformation
\begin{equation}\label{timer}
V_s f(t) = e^{ist}f(t)
\end{equation}
and we will denote the generator of $V_s$ as $\hat t^{r}$. If we think of $L^2(\mathbb{R})$ as the Hilbert space of one {\it extra degree of freedom} living on $\mathbb{R}$ we recognize $H^r$ as the {\it momentum} operator and $\hat t^r$ as the {\it position} operator. Thus mentally we can think of $L^2(\mathbb{R})$ as representing {\it an extra quantum system with just one degree of freedom}. Let us now define a weight $\bar \omega$ on $B(L^2(\mathbb{R}))$ by 
\begin{equation}\label{weight}
\bar \omega (x) = tr(e^{H^r}x)
\end{equation}
with $x\in B(L^2(\mathbb{R}))$ and $tr$ the standard trace on $B(L^2(\mathbb{R}))$. In standard quantum mechanics formal notation we will write $tr(x)=\int_{-\infty}^{+\infty} d\epsilon \langle \epsilon|x|\epsilon\rangle$ with $|\epsilon \rangle$ representing a "momentum" basis of $L^2(\mathbb{R})$ i.e. in our notation $H^r|\epsilon\rangle = \epsilon|\epsilon\rangle$.

Note, as already discussed in the introduction, that (\ref{weight}) has an interesting physical interpretation. If we think of $H^r$ as a formal "Hamiltonian" that in our case is simply the hermitian but not {\it positive} momentum operator in $L^2(\mathbb{R})$, we can think of (\ref{weight}) as formally defining a {\it canonical Gibbs partition function} and of $\bar \omega (x)$ as the corresponding {\it thermal expectation value}\footnote{As we will discuss later this thermality defined by $\bar\omega$ on the {\it extra degree of freedom} Hilbert space $L^2({\mathbb{R}})$ is what is described in \cite{Susskind3} as {\it tomperature}.}. This can be done more explicitely if we think the dimensionless parameter $t$ of $\mathbb{R}$ as $\beta \tilde t$ with $\tilde t$ having units of time. After these preliminaries on $B(L^2(\mathbb{R}))$ we will,
following Connes-Takesaki, consider the algebra $M\otimes B(L^2(\mathbb{R}))$ and define a faithful weight $\phi$ on $M$ and a faithful weight $\omega$ on this algebra as 
\begin{equation}
\omega=\phi\otimes\bar \omega
\end{equation}
Now we can define the transformation $\sigma_{\phi}^{t}\otimes U(t)$ defined by $\omega$. The generator of this transformation is given by
\begin{equation}
h_{\phi}+H^r
\end{equation}
This implies that the {\it Centralizer} of $\omega$ is, by definition, 
\begin{equation}\label{centralizer}
({\cal{M}}\otimes B(L^2(\mathbb{R})))^{(h_{\phi}+H^r)}
\end{equation}
Since by construction $\omega$ satisfies the KMS condition we get that $\omega(x)$ for $x\in{({\cal{M}}\otimes B(L^2(\mathbb{R})))}^{(h_{\phi}+H^r)}$ defines a trace $Tr$. 

To complete the job we need to identify $({\cal{M}}\otimes B(L^2(\mathbb{R})))^{(h_{\phi}+H^r)}$ with a type $II_{\infty}$ factor. This is done once we first identify $({\cal{M}}\otimes B(L^2(\mathbb{R})))^{(h_{\phi}+H^r)}$ with the crossed product of ${\cal{M}}$ by the modular group and we use for ${\cal{M}}$ a type $III_1$ factor.

\subsection{The crossed product}
Let us recall the definition of the crossed product of a type $III_1$ factor by the modular action.
For a type $III_1$ factor ${\cal{M}}$ acting on a Hilbert space $H$ and a normal and faithful state $\phi$ we can define the corresponding GNS representation $\pi_{\phi}$ with $\pi_{\phi}({\cal{A}})$ acting on the GNS Hilbert space $H^{\phi}$ by {\it left multiplication} and with $|1\rangle$ the state in $H^{\phi}$ corresponding to the identity. This, as already discussed,  allows us to define the corresponding modular action $\sigma_{\phi}^{t}$ and the modular hamiltonian $h_{\phi}$ with the normal state $\phi$ satisfies the KMS condition with respect to this modular action. Once you have chosen $\phi$ you can define the crossed product of ${\cal{M}}$ by the modular action $\sigma_{\phi}(t)$,
\begin{equation}
{\cal{M}}^{cr} \equiv {\cal{M}}\rtimes_{\sigma_{\phi}} {\mathbb{R}}
\end{equation} 
The crossed product algebra ${\cal{A}}^{cr}$ is acting on the Hilbert space $L^2(\mathbb{R},H^{\phi})= H^{\phi}\otimes L^2(\mathbb{R})$ \footnote{Hence it is acting on the {\it extra quantum degree of freesdom} with Hilbert space $L^2({\mathbb{R}})$.}. Let us define, following the steps of the former section, the action of time translations on $L^2(\mathbb{R})$ as generated by a "Hamiltonian" operator $H^{r}$, that recall is simply the conjugated momentum for the extra degree of freedom. In these conditions ${\cal{M}}^{cr}$ can be represented as
\begin{equation}
{\cal{M}}^{cr}= {({\cal{M}} \otimes B(L^2(\mathbb{R})))}^{(h_{\phi} +H^{r})}
\end{equation}
where we recognize the centralizer (\ref{centralizer}) of $\omega=\phi\otimes \bar\omega$ for $\bar \omega$ defined in (\ref{weight}). 

Thus ${\cal{M}}^{cr}$ is the desired type $II_{\infty}$ factor on which we define a trace $Tr$ using $\omega$. Thus for elements $\hat a$ in ${\cal{M}}^{cr}$  we get $Tr(\hat a) = \omega(\hat a)$. For $\hat a$ associated to a {\it square integrable} function $a(t)$ valued in ${\cal{M}}$ we get
\begin{equation}\label{basic}
Tr(\hat a) = \omega(\hat a) = \int_{-|infty}^{+\infty} d\epsilon {\cal{F}}( \langle 1| a(t) |1\rangle) e^{\epsilon}
\end{equation}
where ${\cal{F}}$ represents the Fourier transform of $\langle 1| a(t) |1\rangle$ with $|1\rangle$ the state associated to the identity of ${\cal{M}}$ in the GNS construction for the state $\phi$. 

The derivation of (\ref{basic}) can be easily understood. Use $\langle 1| a(t) |1\rangle=: f(t)$ to define a state in $L^2(\mathbb{R})$ let us say the state $|f\rangle$ and the operator $ \hat f=:|f\rangle\langle f|$. Now $\omega(\hat a)= \omega(\hat f)= tr(\hat f e^{H^r})$ that gives you (\ref{basic}). This is the expression used in \cite{Witten1,Witten2,Witten3} for the type $II_{\infty}$ trace.

Summarizing: given a type $III_1$ factor $M$ and a faithful weight $\phi$ on $M$ we define for $M\otimes B(L^2(\mathbb{R}))$ the weight $\omega= \phi\otimes \bar \omega$ with $\bar \omega(x)=tr(e^{H^{r}} x)$. The centralizer ${\cal{C}}(\omega)$ is the crossed product of $M$ by the modular automorphism defined by $\phi$. This is a type $II_{\infty}$ factor with trace defined by $Tr(\hat a)= \omega(\hat a)$ for any $\hat a\in (M\otimes B(L^2(\mathbb{R})))^{(h_{\phi}+H^r)}$.

\subsection{Some additional remarks}
\subsubsection{Uniqueness of $Tr$}
In the previous section  we have observed that the $Tr$ on a type $II_{\infty}$ factor is unique up to rescalings. Thus we should find for $Tr$ defined by $\omega$ on the corresponding centralizer the effect of the rescaling. The rescaling is generated by the action of $V_s$ defined in (\ref{timer}) as
\begin{equation}\label{timeraction}
(1\otimes V_s)\omega= e^{s}\omega
\end{equation}
i.e. the rescaling $Tr\rightarrow \alpha Tr$ is induced by $V_s$ for $\alpha=e^s$. Recall that the generator of $V_s$ is the conjugated of $H^r$ that we denoted as $\hat t^{r}$.

\subsubsection{Thermality and $Tr$}
One remarkable fact of the trace $Tr$ defined by $\omega$ is the apparent thermality encoded in the weight $\bar \omega$ relative to $H^r$. The meaning of this formal thermality defined for the quantum system described by $L^2(\mathbb{R})$ will become clear in concrete examples once we identify the physical meaning of the {\it auxiliary quantum system} defined by the Hilbert space $L^2(\mathbb{R})$.

\subsubsection{The Hilbert space picture}
Another remarkable fact is that the Hilbert space on which the crossed product algebra  ${\cal{M}}^{cr}$ is acting is $H\otimes L^2(\mathbb{R})$ with $H$ the Hilbert space on which the type $III_1$ factor ${\cal{M}}$ is defined. This as stressed in \cite{Witten2} clashes with our standard quantum field theory intuition. Indeed ${\cal{M}}^{cr}$ represents operators commuting with $h_{\phi}+H^r$ so we should expect the Hilbert space to be defined by the subspace of states $|\psi\rangle$ in $H\otimes L^2(\mathbb{R})$ satisfying the constraint $(h_\phi+H^r)|\psi\rangle=0$. This will be the case when the constraint represents the invariance under the action of a compact group. In our case we observe that the "physical Hilbert space" representing states "created" by elements in ${\cal{M}}^{cr}$, formally invariant under the reparametrizations generated by $h_{\phi}+H^r$, is precisely the space $H\otimes L^2(\mathbb{R})$ i.e. the space representing the crossed product.

\subsubsection{The type $II_1$ projection}
As already discussed for a projection $p\in {\cal{M}}^{cr}$ with finite $Tr(p)$ we can define the type $II_1$ factor $p{\cal{M}}^{cr}p$. On this type $II_1$ factor we have a unique and well defined normalizable trace $tr$. For a projection $p$ we can define $pH^rp$ and $p\mathbb{R}$ the spectrum of $pH^rp$. In these conditions the trace $tr$ on the type $II_1$ factor can be obtained by projecting equation (\ref{basic}), namely
\begin{equation}\label{basicc}
tr_p(\hat a)= \int_{p\mathbb{R}} d\epsilon {\cal{F}}(\langle 1|a(t)|1\rangle) e^{\epsilon}
\end{equation}
and requiring the integral in (\ref{basicc}) to be finite.

\subsubsection{The microcanonical TFD and the maximal entropy state}
In $L^2({\mathbb{R}},H^{\phi})$ we can consider some particular {\it product states} $|1,f\rangle =: |1\rangle. f$ with $f\in L^2(\mathbb{R})$ and with $\int|f|^2=1$. The density matrix $\rho_{|1,f\rangle}$ is defined by
\begin{equation}
Tr(\rho_{|1,f\rangle} \hat a)= \langle 1,f|\hat a|1,f\rangle
\end{equation}
This implies
\begin{equation}
\bar \omega(\rho_{|1,f\rangle}) = |f|^2
\end{equation}
for $\bar \omega$ defined in (\ref{weight}) i.e. $\rho_{|1,f\rangle}= e^{-H^r} |f|^2$. For the two sided black hole metric and for $|1\rangle$ the Hartle Hawking state, the state $|1,f\rangle$ is the microcanonical TFD state used in \cite{Witten3}. 

Let us now consider the type $II_1$ factor obtained after a projection $p$. The product state in $pL^2(\mathbb{R},H)$ corresponding to maximal entropy $\rho_{max}=1$ is the state $|1\rangle .f_{max}$, with $|1\rangle$ the corresponding GNS representative of the identity, and with $f_{max}$ such that
\begin{equation}\label{thermal}
\bar \omega_p(1) = tr(e^{pH^rp}) = |f_{max}|^2
\end{equation}
which for the spectrum of $pH^{r}p$ equal $[0,+\infty]$ implies $f_{max}=e^{-\frac{\epsilon}{2}}$ with $\int_0^{+\infty}|f_{max}|^2 =1$. In summary for the type $II_{\infty}$ factor ${\cal{M}}^{cr}$ we can select the {\it microcanonical states} $|1\rangle.f$ with $f\in L^2(\mathbb{R})$ while for the type $II_1$ factor $p{\cal{M}}^{cr}p$ we can select the {\it canonical} maximal entropy state $|1\rangle.e^{- \frac{\epsilon}{2}}$. The physical meaning of these states will be discussed in section 4 \footnote{Note that the thermality of $\bar\omega$ in (\ref{thermal}) i.e.{\it the tomperature} \cite{Susskind3} is manifest in the form of $f_{max}$.}.

\subsubsection{Connection to {\it integrable time}}
Finally let us recall from the introduction that the space $L^2({\mathbb{R}},H^{\phi})$ where we have defined the action of the centralizer can be interpreted as the space of {\it time histories} $|\psi\rangle(t)$ in $H^{\phi}$ ( or more generically in $H_{QFT}$ ). Thus, as stressed in the introduction the emergence of the extra quantum degree of freedom is the consequence of defining {\it integrable time transformations} on $H_{QFT}$. It is this integrability, namely that $\int|||\psi\rangle (t)||^2$ is finite what naturally delivers the existence of an extra degree of freedom\footnote{As mentioned in the introduction it is tempting to associate the notion of integrable time with some form of {\it complexity} measured by $\int|||\psi\rangle (t)||^2$ and to connect finite complexity with the emergence of the extra ( observer ) degree of freedom.}
\section{Clocks and the extra degrees of freedom}

\subsection{Observers in the weak gravity limit}
From now on our basic framework will be QFT defined on a fixed space-time background coupled to gravity. Coupling to gravity is achieved by imposing {\it general covariance} which implies that the algebra of {\it physical observables} is defined by those invariant under general coordinate transformations. In what follows we will work in the weak gravity limit. In this limit the only gravitational quantum fluctuations will be those describing linearized gravity on our fixed space-time background. Moreover we will focus on invariance under {\it coordinate time reparametrizations}.

As already said by an {\it observer}, in the space-time background, we will mean a timelike world line parametrized by a real coordinate time $t\in [+\infty,-\infty]$ with a {\it causal domain} ${\cal{D}}$. In addition we will think the observer as an external quantum mechanical system with Hilbert space $H_{obs}$ and with an algebra of observables ${\cal{A}}_{obs}$ acting on $H_{obs}$ describing the observer properties. The minimal definition of the observer will be: $H_{obs}=L^2(\mathbb{R})$ and ${\cal{A}}_{obs}= B(L^2(\mathbb{R}))$ with coordinate time $t$ parametrizing $\mathbb{R}$. In addition to the observer we have the QFT Hilbert space $H_{QFT}$ as well as the algebra ${\cal{A}}({\cal{D}})$ of the QFT local observables the observer can measure performing local measurements in the causal domain ${\cal{D}}$. Thus the formal frame where we will work is characterized by a Hilbert space 
\begin{equation}\label{Hspace}
{\cal{H}} =: H_{QFT}\otimes L^2(\mathbb{R})
\end{equation} 
and an algebra of operators ${\cal{A}}({\cal{D}}) \otimes B(L^2(\mathbb{R}))$ acting on ${\cal{H}}$.

The group $G$ of time reparametrizations is the non compact group of time translations on $\mathbb{R}$. For an {\it ideal observer} the action of $G$ on ${\cal{A}}({\cal{D}}) \otimes B(L^2(\mathbb{R}))$ will be defined by some $\bar \alpha(t)\in Aut ({\cal{A}}({\cal{D}}) \otimes B(L^2(\mathbb{R})))$ of the form
\begin{equation}
\bar \alpha(t) = \alpha(t) \otimes \lambda(t)
\end{equation}
with $\alpha(t) \in Aut ({\cal{A}}({\cal{D}}))$ and with $\lambda(t)\in Aut(B(L^2(\mathbb{R}))$ defined by the unitary transformation $U(t_0)f(t)=f(t+t_0)$ for $f\in L^2(\mathbb{R})$. In the previous section we denoted the generator of $U(t)$ as $H^r$ with $H^r=-i\hbar\frac{d}{dt}$. Here it would be convenient to denote this generator as $H^{obs}$ \footnote{In this context we are thinking the extra degree of freedom discussed in previous sections as representing an external observer. This is just a nice physical interpretation, however the important think, as already stressed, is the addition of an extra quantum degree of freedom.}.

The principle of {\it general covariance} allows us to define the algebra of physical observables ${\cal{A}}_{phys}({\cal{D}})$ as the elements in ${\cal{A}}({\cal{D}}) \otimes B(L^2(\mathbb{R}))$ {\it invariant under the action of $\bar \alpha(t)$}. Note that at this level of the discussion the {\it observer} simply represents an {\it extra and external degree of freedom} that we identify with the coordinate time $t$. Moreover the representation of the Hilbert space in (\ref{Hspace}) is that of $L^2({\mathbb{R}},H_{QFT})$ i.e. the space of integrable ( square summable) quantum paths $|\psi|rangle(t)$ in $H_{QFT}$.

Let us now  {\it assume} that ${\cal{A}}({\cal{D}})$ is a von Neumann algebra and that we define $\alpha(t)$ as {\it the modular automorphism} $\alpha_{\phi}(t)$ defined by a {\it normal state} $\phi$ on ${\cal{A}}({\cal{D}})$. We can also look for a weight $\omega$  on ${\cal{A}}({\cal{D}}) \otimes B(L^2(\mathbb{R}))$ such that the corresponding modular transformation  is $\alpha_{\omega}(t)=\alpha_{\phi}(t) \otimes \lambda(t)$. Using Connes Takesaki theorem
we get
\begin{equation}
\omega=\phi\otimes \bar \omega
\end{equation}
with $\bar\omega(x)= tr(xe^{H^{obs}})$. Once this is done we define the physical algebra
\begin{equation}
{\cal{A}}_{phys}({\cal{D}}) = {\cal{C}}(\omega)
\end{equation}
i.e. the centralizer of the weight $\omega$.

As discussed in the previous section this is the crossed product algebra
\begin{equation}
{\cal{A}}({\cal{D}})\rtimes_{\alpha_{\omega}} {\mathbb{R}}
\end{equation}

\subsection{Thermality and Observers}
One of the key and deeper aspects of this formal way to define the algebra of {\it physical observables} is its profound {\it thermodynamic} flavor. Indeed, as already stressed, the {\it observer contribution} defined by $\bar \omega$ is the standard way to define {\it thermal correlators} for operators in $B(L^2(\mathbb{R})$ using $H^{obs}$ as {\it formal Hamiltonian}. In reality $H^{obs}$ is not a real Hamiltonian with positive spectrum. At this level it is simply the generator representing coordinate time translations on the observer Hilbert space $L^2(\mathbb{R})$ i.e. it is simply the hermitian but {\it not positive} momentum operator. This deep connection can be put in the form of a slogan as:
\begin{center}
{\it General covariance is observer ( extra degree of freedom) formal ( since $H^{obs}$ is at this level not positive ) thermality}
\end{center}
What we want to stress with the {\it slogan} is that the algebraic way to implement general covariance i.e. the definition of ${\cal{A}}_{phys}({\cal{D}})$ as a non trivial centralizer, is using the formally thermal weight $\bar\omega$ on the {\it observer} algebra $B(L^2(\mathbb{R}))$ \footnote{ In the idiolect introduced in \cite{Susskind3} we could say the {\it general covariance} requires {\it finite tomperature}.}. 

The thermality on which we should be interested is in the one associated to $\omega=\phi\otimes \bar\omega$. In order to unveil this thermality we need first to know what type of factor is ${\cal{A}}_{phys}({\cal{D}})$. If the observer causal domain ${\cal{D}}$ is a {\it bounded region} of space-time we expect from {\it locality} that the QFT algebra ${\cal{A}}({\cal{D}})$ is a type $III_1$ factor which implies that ${\cal{A}}_{phys}({\cal{D}})$ is a type $II_{\infty}$ factor. As already discussed in section 2, this factor is equipped with a trace $Tr$ defined by
\begin{equation}
Tr(\hat a) = \omega(\hat a)
\end{equation}
for $\hat a$ any element in ${\cal{A}}_{phys}({\cal{D}})$. This type $II_{\infty}$ trace is not a finite trace, however we could, in order to unveil the underlying thermality, to define formally the {\it entropy} of the {\it ground state} as
\begin{equation}
S= \ln Tr(1)
\end{equation}
for $1$ the identity in ${\cal{A}}_{phys}({\cal{D}})$. This identifies the "ground state" with the GNS state defined by $\phi$ associated to the identity. This formal type $II_{\infty}$ entropy is then determined by the observer $\bar\omega$ as
\begin{equation}
S=\ln \bar \omega(1)=\ln tr(e^{H_{obs}})
\end{equation}
that is what we have identified as {\it observer thermality}. 

As already discussed in section 2, the action of $V_s$ defined in (\ref{timeraction}) changes $S$ as
\begin{equation}
S\rightarrow S+s
\end{equation}
An appealing aspect of the former construction is that the potential dependence on $\phi$ is fully under control thanks to Connes cocycle condition which implies independence on the chosen normal state $\phi$ on ${\cal{A}}({\cal{D}})$. Recall that for other states $|\Psi\rangle$ different from the "ground state", defined by the GNS construction for $\phi$, we can define the associated density matrix operator $\rho_{|\Psi\rangle}$ by
\begin{equation}
\langle \Psi|\hat a|\Psi\rangle = Tr(\rho_{|\Psi\rangle} \hat a) = \omega(\rho_{|\Psi\rangle} \hat a)
\end{equation}
and to define the entropy associated to $|\Psi\rangle$ as
\begin{equation}
S(|\Psi\rangle)= Tr(\ln \rho_{|\Psi\rangle}) = \omega(\ln \rho_{|\Psi\rangle})
\end{equation}
\subsection{The effect of projections}
For ${\cal{A}}_{phys}({\cal{D}})$ a type $II_{\infty}$ factor we can define a type $II_1$ factor as $p{\cal{A}}_{phys}({\cal{D}})p$ for a projection $p$ with $Tr(p)$ finite. The $tr$ on this type $II_1$ factor is defined as
\begin{equation}
tr(\hat x)= \omega_p(\hat x)
\end{equation}
for any $\hat x \in p{\cal{A}}_{phys}({\cal{D}})p$ and with 
\begin{equation}
\omega_p =\phi \otimes \bar \omega_p
\end{equation}
with $\bar \omega_p$ defined by $pH^{obs}p$. 

Now while $Tr$ is {\it unique}, up to rescalings defined by $V_s(Tr)= e^s Tr$, the type $II_1$ trace $tr$ defined above is unique. So {\it formally} we could try to represent $tr$ in terms of $Tr$ as a formal integral over the scale parameter i.e.
$tr= \int ds \mu(s) V_s(Tr)$ for some measure over the scale parameter. Obviously this construction has only a purely heuristic meaning intended to identify for the {\it scale} transformations defined by $V_s$ the trace $tr$ of the type $II_1$ factor as {\it scale invariant} or in more physical terms, as we will see in a moment, as {\it breaking the scale flow of weights transformation}.

Note that the generator of $V_s$, that is the operator $\hat t_{obs}$ conjugated to $H^{obs}$ introduced in (\ref{timer}), is {\it not} part of ${\cal{A}}_{phys}({\cal{D}})$, actually is an outer automorphism, although it is obviously in ${\cal{A}}({\cal{D}})\otimes B(L^2(\mathbb{R}))$. We could try to include $\hat t_{obs}$ by defining the crossed product of ${\cal{A}}_{phys}({\cal{D}})$ by the action of $V_s$. This {\it double crossed product} is by {\it Takesaki duality} isomorphic to ${\cal{A}}({\cal{D}})\otimes B(L^2(\mathbb{R}))$ \footnote{Note that although the trace $tr$ defined for the projected type $II_1$ factor leads to {\it flat entanglement} it keeps finite the {\it tomperature} defining the $\bar \omega$ weight.}.

\subsubsection{The flow of weights}
The flow of weights in this case is identified with the action of $V_s$ on the central elements of the centralizer i.e. on ${\cal{Z}}({\cal{A}}_{phys}({\cal{D}}))$ with ${\cal{A}}({\cal{D}})$ a type $III_1$ factor. In this case the flow of weigths reduces to shifts of $H^{obs}$ by a constant $H^{obs}\rightarrow H^{obs}+s$, namely
\begin{equation}
V_s \bar \omega = e^{s} \bar \omega
\end{equation}
In this sense in the type $II_{\infty}$ centralizer this flow of weights acts as an automorphism. We will describe this situation saying that in the type $II_{\infty}$ regime the flow of weights defines a {\it scale symmetry transformation} on the weights. In more physical terms we can think of this transformation as changing the background defined by $\omega=\phi\otimes \bar\omega$.

Note now that using a finite projection $p$ to define a type $II_1$ factor we effectively:

{\it Break the symmetry under the transformation defining the flow of weights} 

\subsection{Making the observer a physical clock}
Until now we have defined the observer using the algebra $[H^{obs},\hat t_{obs}]=i\hbar$ acting on $L^2(\mathbb{R})$ as standard {\it position and momentum} operators. The reason to use
the notation $H^{obs}$ and to think this operator as some Hamiltonian is because we use coordinate time $t$ to define $\mathbb{R}$ in $L^2(\mathbb{R})$ and consequently the "momentum" operator $H^{obs}$ defines "time translations". 

However we could, and probably we should, be interested in interpreting the observer as a {\it real physical clock}. How that can be done? 

We should think in a quantum mechanical model defined by a generic {\it position and momentum Heisenberg algebra}, let us say $p_{clock},q_{qlock}$ with $[p_{clock},q_{qlock}]=i\hbar$ and with some {\it clock dynamics} defined by an {\it hermitian and positive} Hamiltonian $H_{clock}$ in terms of which we define both $\dot p_{clock}$ and $\dot q_{clock}$ using the standard Heisenberg equations. 

Now we can define a formal clock time hermitian operator $\hat t_{clock}$ satisfying
\begin{equation}
[H_{clock},\hat t_{clock}]= i\hbar
\end{equation}
Following \cite{ABc}\cite{Gomez3} we can define, for instance, 
\begin{equation}\label{clocktime}
\hat t_{clock}\sim q_{clock}\frac{1}{p_{clock}} + \frac{1}{p_{clock}} q_{clock}
\end{equation}
and a free clock Hamiltonian $H_{clock}=p_{clock}^2$
provided {\it we work on states in the clock Hilbert space $L^2(\mathbb{R})$ ( where now $\mathbb{R}$ represents the spectrum of the clock position operator $q_{clock}$ ) such that $\langle p_{clock}\rangle$ is non vanishing and with {\it small variance} of $\dot q_{clock}$}. Let us formally define $\tilde L^2(\mathbb{R})$ as the states in $L^2(\mathbb{R})$ for which $\langle p_{clock}\rangle$ is different from zero and such that the variance of $\dot q_{clock}$ is very small i.e. as the space of {\it good clock states}. Obviously this is not a rigorous definition of $\tilde L^2(\mathbb{R})$, but let us ignore for the time being this problem. 

Now the observer, as a real clock, is defined for some $H_{clock}$ by the algebra generated by $H_{clock},\hat t_{clock}$ and we can now try to define ${\cal{A}}_{phys}({\cal{D}})$ as a crossed product but formally using $\tilde L^2(\mathbb{R})$ and the algebra $H_{clock},\hat t_{clock}$ instead of $L^{2}(\mathbb{R})$ and $H^{obs},\hat t_{obs}$. 

More precisely the traces for this algebra will be defined by
\begin{equation}
\omega_{clock}=\phi\otimes \bar \omega_{clock}
\end{equation}
with $\bar \omega_{clock}$ defined as $\bar \omega_{clock}(x)=tr(xe^{H_{clock}})$ \footnote{With this trace $tr$ properly defined on the set of good clock states.}. Elements in the so defined ${\cal{A}}_{phys}({\cal{D}})$ are generated by $e^{i\hat t_{clock} h_{\phi}}ae^{-i\hat t_{clock}h_{\phi}}$\footnote{These are the {\it clock dressed} operators used in \cite{Gomez3}.}.

Using the previous definitions we notice that what we have called before an observer is just a clock with $p^{clock}=H^r$ and $q^{clock}= t^r$ but with vanishing  hermitian and positive clock Hamiltonian. When we work with a real clock with positive and  hermitian $H_{clock}$ the so defined crossed product algebra should be formally related to a type $II_1$ factor defined by a projection $p$ such that $pH^{obs}p\sim H_{clock}$. In that sense we observe that:
\begin{center}
{\it the clock breaks the invariance under the flow of weights}
\end{center}
Indeed once we move into the effective clock algebra defined by $H_{clock}$ and $t_{clock}$ the formal role of $t_{clock}$ as generating flows of weights disappears.

\subsection{Hawking clock}
For reasons that will become clear in the next section we will call Hawking clock a clock with non trivial clock Hamiltonian $H_{clock}$ inducing non vanishing $\dot p_{clock}$. As we will discuss in the next section, where we will consider concrete examples of the former mathematical construction, for the case of the two sided black hole we can identify $H^{obs}$ as the ADM mass $M_{ADM}$. When we promote this observer into a physical clock, identifying $H^{obs}=p_{clock}$, the introduction of a  non trivial $H_{clock}$ with non vanishing $\dot p_{clock}$ implies a non vanishing $\dot M_{ADM}$. A clock time operator can be then defined using (\ref{clocktime}) as
\begin{equation}\label{hawking}
\hat t_{clock} \sim \frac{M_{ADM}}{\langle \dot M_{ADM} \rangle}
\end{equation}
making very explicit the origin of the name Hawking clock. Interestingly enough $\dot M_{ADM}$ is a {\it quantum gravity effect} due to the gravitational back reaction of Hawking emission. In this sense we can suggest that {\it the black hole algebra defined relative to a Hawking clock} is a type $II_1$ factor and that the transformation, in this case of type $II_{\infty}$ into a type $II_1$ is due to take into account a purely gravitational back reaction effect physically reflecting {\it the quantum breaking of the flow of weights symmetry}. This rough comment  induces us to modify the former slogan into

{\it Gravity ( in the sense of non vanishing $\dot M_{ADM}$ ) is clock thermality ( encoded in the clock weight $\bar\omega_{clock}(x)= tr(xe^{H_{clock}})$ )}

or, equivalently, although probably making a too free use of language: {\it Gravity emerges as a consequence of the breaking of the flow of weights symmetry.}
In the next section we will briefly develop these ideas.

\section{de Sitter and Black Holes in correspondence}

In this final section we will consider de Sitter (dS) and black holes (BH) in parallel. Let us define ${\cal{A}}_{dS}$ and ${\cal{A}}_{BH}$ the two type $III_1$ factors acting on the QFT Hilbert spaces $H_{dS}$ and $H_{BH}$. These factors are interpreted as the algebras of local operators with support on the dS static patch or on the external region of the two sided black hole.

Let us define two positive and faithful normal states $\phi_{dS}$ and $\phi_{BH}$ and define the corresponding GNS Hilbert spaces $H^{\phi_{dS}}$ and $H^{\phi_{BH}}$. These Hilbert spaces will be the basic arena to define QFT on these classical backgrounds. Since both $\phi_{dS}$ and $\phi_{BH}$ are not satisfying the trace property both will induce, by lifting the adjoint action to $H^{\phi}$, the corresponding modular automorphisms $\sigma^t_{dS}$ and $\sigma^t_{BH}$ where we ignore for notational simplicity the explicit reference to $\phi_{dS}$ and $\phi_{BH}$. Fortunately this {\it background dependence} is under control due to Connes cocycle condition. The GNS ground states are the states in $H^{\phi}$ associated to the identity in ${\cal{A}}_{dS}$ and ${\cal{A}}_{BH}$ and will be denoted $|1_{dS}\rangle$ and $|1_{BH}\rangle$. 

Once we count with the GNS representation we can define the commutant ${\cal{A}}_{dS}^{'}$ and ${\cal{A}}_{BH}^{'}$ relative to the action of these algebras on the GNS Hilbert spaces. Let us denote $h_{dS}$ and $h_{BH}$ the modular generators of $\sigma^t_{dS}$ and $\sigma^t_{BH}$ respectively. Since by construction $|1_{dS}\rangle$ and $|1_{BH}\rangle$ are cyclic and separating we can define the spectrum of $h_{dS}$ and $h_{BH}$ on states in the clousure of ${\cal{A}}_{dS}|1_{dS}\rangle$ and ${\cal{A}}_{BH}|1_{BH}\rangle$. In both cases this spectrum is $[0,+\infty]$. Since in both cases $H^{\phi}$ describes the QFT Hilbert space on the whole Penrose diagram we can choose $|1_{dS}\rangle$ and $|1_{BH}\rangle$ to be invariant under the full group $G$ of isometries. Note that only a subgroup of $G$ is acting on the static patch or on the half sided black hole region.

Let us now define an {\it integrable time} or equivalently {\it let us add an extra degree of freedom}. As discussed in previous sections this means to define the algebras ${\cal{A}}_{dS}\otimes B(L^2(\mathbb{R}_{dS}))$ and ${\cal{A}}_{BH}\otimes B(L^2(\mathbb{R}_{BH}))$ with ${\mathbb{R}}_{dS,BH}$ two identical copies of $\mathbb{R}$ and with two unitary translation operators $U_{dS}(t_0)$ and $U_{BH}(t_0)$ acting on $L^2(\mathbb{R})$ as time translations $U(t_0)f(t)=f(t+t_0)$. Let us denote, using the discussion in the previous section, the corresponding generators as $H^{obs}_{dS}$ and $H^{obs}_{BH}$ and let us define the two faithful weights
\begin{equation}
\omega_{dS,BH}=\phi_{dS,BH}\otimes \bar\omega_{dS,BH}
\end{equation}
with, as described in the previous section, $\bar\omega_{dS,BH}(x)=tr(xe^{H^{obs}_{dS,BH}})$.

Using $\omega_{dS,BH}$ we can define the modular automorphisms $\sigma^t_{\omega_{dS,BH}}$ with modular generators $h_{dS,BH}+H^{obs}_{dS,BH}$. The {\it centralizers} ${\cal{C}}(\omega_{dS,BH})$ are
\begin{equation}\label{centralizerrepre}
{\cal{C}}(\omega_{dS,BH})= ({\cal{A}}_{dS,BH}\otimes B(L^2(\mathbb{R}_{dS,BH})))^{(h_{dS,BH}+H^{obs}_{dS,BH})}
\end{equation}
that are both type $II_{\infty}$ factors equipped with a trace $Tr$ defined by
\begin{equation}
Tr_{dS,BH}(a)=\omega_{dS,BH}(a)
\end{equation}
for $a\in {\cal{C}}(\omega_{dS,BH})$.

In both cases this $Tr$ is unique up to rescalings $Tr\rightarrow e^sTr$ generated by $V^{dS}_s$ and $V^{BH}_s$ generated by $t^{obs}_{dS,BH}$ the operator canonically conjugated to $H^{obs}_{dS,BH}$. Note that at this level of the discussion $H^{obs}$ and $t^{obs}$ are just the standard momentum and coordinate operators, associated to the extra degree of freedom with Hilbert space $L^2(\mathbb{R})$ \footnote{It is important to keep in mind that in the former construction we start, in both cases, with the physics data ${\cal{A}}_{dS,BH}$ acting on $H_{dS,BH}$ and with the normal states $\phi_{dS,BH}$. Recall that normal states are vector states in the extended Hilbert space $H_{dS,BH}\otimes l^2(N)$ with $\phi_{BH,dS}(x) =\sum\langle \psi_i|x|\psi_i\rangle$ and $\psi_i\in l^2(N,H_{BH,dS})$. Thus we have two extensions of the Hilbert space $H_{dS,BH}$. One in order to define the normal states as vector states and another in order to define the integrable time.}

\subsection{The type $II_{\infty}$ Black Hole case and the flow of weights.}
This type $II_{\infty}$ factor is defined by the centralizer ${\cal{C}}(\omega_{BH})$. As discussed this allows us to define a trace $Tr_{BH}$ on ${\cal{C}}(\omega_{BH})$ as
\begin{equation}
Tr_{BH}(a) = \omega_{BH}(a)
\end{equation}
for any $a \in {\cal{C}}(\omega_{BH})$. This trace $Tr_{BH}$ is not normalizable with
\begin{equation}
Tr_{BH}(1)=tr(e^{H^{obs}_{BH}}) = \infty
\end{equation}
Note that although the modular Hamiltonian $h_{\phi_{BH}}$ is not part of the type $III_1$ factor ${\cal{A}}_{BH}$ the operators $h_{\phi_{BH}}$ and $H^{obs}_{BH}$ {\it are in ${\cal{C}}(\omega_{BH})$}. Thus we can define the operator
\begin{equation}\label{extra}
H_{BH}^{'} =: h_{\phi_{BH}} + H^{obs}_{BH}
\end{equation}
and consequently we can get {\it once we work with the type $II_{\infty}$ factor ${\cal{C}}(\omega_{BH})$} the representation of $h_{\phi_{BH}}$ as
\begin{equation}\label{decomposition}
h_{\phi_{BH}} = H_{BH}^{'}-H^{obs}_{BH}
\end{equation}
However using (\ref{centralizerrepre}) and since $h_{\phi_{BH}}$ is not part of ${\cal{A}}_{BH}$ we can conclude that the operator defined in (\ref{extra}) is associated with the extra degree of freedom\footnote{More precisely is in the commutant of ${\cal{C}}(\omega_{BH})$.}. In this case the expression (\ref{decomposition}) acquires a nice and deep meaning. Namely we can think of $H_{BH}^{'}$ as the result of {\it shifting} by some automorphism $V_s$, defining the {\it flow of weights}, the Hamiltonian $H^{obs}_{BH}$. 

In summary:

{\it For ${\cal{C}}(\omega_{BH})$ the modular Hamiltonian $h_{\phi_{BH}}$ reflects the action generated by the flow of weights.}

Recall that adding the extra degree of freedom, to which we are referring here as an observer, is the way to introduce what we described already in the introduction as {\it an integrable time} i.e. square summable functions $|\Psi\rangle(t)$ in the Hilbert space $H^{\phi_{BH}}$. The extra degree of freedom reflects the natural isomorphism of $L^2(\mathbb{R},H^{\phi})=H^{\phi}\otimes L^2(\mathbb{R})$. Note that for the two sided black hole (\ref{decomposition}) represents the {\it time shift} introduced in \cite{Witten3}. 

Equation (\ref{decomposition}) is also specially nice to understand physically the effect of moving into a type $II_1$ description using a finite projection $p$ in ${\cal{C}}(\omega_{BH})$. As discussed $V_s$ is not commuting with the projection implying the vanishing of the time shift. 

\subsection{The type $II_{\infty}$ dS}
For dS we can, following exactly the same steps, to define the type $II_{\infty}$ centralizer ${\cal{C}}(\omega_{dS})$ as well as $Tr_{dS}$ with $Tr_{dS}(1)= tr(e^{H^{obs}_{dS}}) = \infty$. In other words in this $II_{\infty}$ version we don't have maximal entropy state. 

The obvious question at this point is : Why in the case of dS we move into the type $II_1$ version? The physics reason is that in the case of dS the action of the flow of weights $V_s$ is defining a {\it gauge transformation} that we should mod out in order to define the algebra of physical observables \footnote{As explained in \cite{Witten2} the reason for that is that in the de Sitter case the action of $V_s$ generates pure gauge fluctuations.}.

This means that for dS you must consider the $H^{'}$ and $H_{dS}^{obs}$ in (\ref{decomposition}) as {\it gauge equivalent} or equivalently that  $h_{\phi}$ defined by (\ref{decomposition}) is gauge equivalent to zero which is indeed the definition of the trivial modular dynamics for type $II_1$ i.e. of {\it flat entanglement}.

\subsection{The type $II_1$ Black hole: Hawking clock}
Let us now promote the extra degree of freedom used to define the type $II_{\infty}$ centralizer ${\cal{C}}(\omega_{BH})$ into a physical clock. This, as discussed in the previous section, requires to define, for the extra degree of freedom with coordinate and momentum operators $t^{obs}$ and $H^{obs}$, an hermitian and {\it positive} operator $H^{clock}$ with $[H^{clock},H^{obs}]=\dot H^{obs}$ and $[H^{clock},t^{obs}]=\dot t^{obs}$. Now let us define ${\cal{H}}^{clock}$ the elements in $L^2(\mathbb{R})$ for which $\langle \dot H^{obs} \rangle$ is non vanishing and for which the variance of $\dot H^{obs}$ is sufficiently small i.e. the space of good clock states. 

In what follows we will not need to make explicit $H^{clock}$ but just the typical value of $\dot H^{clock}$.

Once we have defined $H^{clock}$ and ${\cal{H}}^{clock}$ we can define the clock weight
\begin{equation}
\omega_{clock}=\phi_{BH}\otimes \bar \omega_{clock}
\end{equation}
with 
\begin{equation}
\bar \omega_{clock} (x)= tr_{{\cal{H}}^{clock}}(e^{H^{clock}}x)
\end{equation}
for any $x\in B(L^2(\mathbb{R}))$.

We will assume that:

{\it The centralizer ${\cal{C}}(\omega_{clock})$ is a type $II_1$ factor}

with the elements in ${\cal{C}}(\omega_{clock})$ {\it clock dressed} using as dressing $e^{i\hat t^{clock} h{\phi}} a e^{-i\hat t^{clock} h{\phi}}$ for $ t^{clock} \sim \frac{t^{obs}}{\langle \dot t^{obs} \rangle}$ on ${\cal{H}}^{clock}$. In other words we use on ${\cal{H}}^{clock}$ the operators $t^{clock}$ and the {\it positive} $H^{clock}$ Hamiltonian to define the type $II_1$ factor. As discussed in the previous section we can always formally define a type $II_1$ factor for the BH simply using a finite projection $p$ and defining the type $II_1$ centralizer as $p{\cal{C}}(\omega_{BH})p$. In that way we could think that there exist a projection $p$ such that
${\cal{C}}(\omega_{clock})
\sim p{\cal{C}}(\omega_{BH})p$ with $\sim$ representing unitary equivalence.

We can define a "dual" version where we interchange the roles of $t^{obs}$ and $H^{obs}$ and where we define $ t^{clock} \sim \frac{H^{obs}}{\langle \dot H^{obs} \rangle}$. It is interesting now to use for the BH the interpretation of $H^{obs}$ as the ADM mass $M_{ADM}$ that leads to expression (\ref{hawking}). Using Stephan Boltzmann law we get
\begin{equation}
\dot M_{ADM} \sim AT_{H}^4
\end{equation}
for $A$ the black hole horizon area  and $T_{H}$ Hawking temperature. Note that:

{\it In the weak gravity limit $\dot M_{ADM}=0$ and consequently we cannot define the type $II_1$ version.}

\subsubsection{The action of the flow of weights}
Now we should ask how the type $II_{\infty}$ flow of weights is acting on $\omega_{clock}$. The answer is physically nice and intuitive. Indeed since the flow of weights is shifting $M_{ADM}$ by a constant, we observe that the {\it free clock Hamiltonian} $H^{clock}\sim (\dot M_{ADM})^2$ remains {\it invariant under the flow of weights}. This is also the case for $\omega_{clock}$ reflecting the uniqueness of the type $II_1$ trace $tr_{clock}$ defined by $\omega_{clock}$. In other words we can think ( probably very generically) the clock
as {\it invariant under the scale flow of weights}.

\subsection{The {\it coupling constant} of the BH type $II_1$ factor and a type $II_1$ Page curve}
Once we have defined the type $II_1$ factor as the centralizer ${\cal{C}}(\omega_{clock})$ we can wonder about the Murray von Neumann dimension ( coupling constant) for this type $II_1$ factor. This factor was defined as acting on $H^{\phi}\otimes {\cal{H}}^{clock}$, thus to define the corresponding coupling constant, let us say $d_{BH}$ we need to define an isometry (\ref{isometry})
\begin{equation}\label{BHisometry}
u_{BH}: H^{\phi}\otimes {\cal{H}}^{clock} \rightarrow H_{tr}\otimes L^2({\mathbb{N}})
\end{equation}
with $H_{tr}$ the GNS representation of the type $II_1$ factor defined by its unique trace $tr$. Then
\begin{equation}
d_{{\cal{A}}_{BH}} = tr(u_{BH}^{+}u_{BH})
\end{equation}
At this point we can only offer a suggestion on this value. For the type $II_1$ factor ${\cal{A}}_{BH}$ defined by the effective projection associated to a clock characterized by some non vanishing $\dot M_{ADM}$ we could define a function $d_{{\cal{A}}_{BH}}(\dot M_{ADM})$, where the dependence on $M_{ADM}$ is implicit in the dependence of $u_{BH}$ on the clock space ${\cal{H}}^{clock}$. For $\dot M_{ADM}=0$ corresponding to a vanishing {\it clock Hamiltonian} we expect ${\cal{A}}_{BH}$ defined as the centralizer ${\cal{C}}(\omega_{BH})$ {\it to be the standard type $II_{\infty}$ factor we normally associate with the BH}. This means (see section 2.4.2) that 
\begin{equation}
d_{{\cal{A}}^{'}_{BH}}(\dot M_{ADM}=0) = \infty
\end{equation}
and consequently assuming $d_{{\cal{A}}^{'}_{BH}}(\dot M_{ADM}) = \frac{1}{d_{{\cal{A}}_{BH}}(\dot M_{ADM})}$ that 
\begin{equation}
d_{{\cal{A}}_{BH}}(\dot M_{ADM}=0) =0
\end{equation}
In this way we can define the {\it curve} $d_{{\cal{A}}_{BH}}(\dot M_{ADM})$ depending on $\dot M_{ADM}$ starting at zero for $\dot M_{ADM}=0$ and growing as a function of $\dot M_{ADM}$ as
\begin{equation}
d_{{\cal{A}}_{BH}}(\dot M_{ADM}) \sim \dot M_{ADM}
\end{equation}
Similarly we can define the {\it dual curve} defined by $d_{{\cal{A}}^{'}_{BH}}(\dot M_{ADM})$ that we expect to be $+\infty$ for $\dot M_{ADM}=0$ and to vanish for $\dot M_{ADM}=\infty$ as $d_{{\cal{A}}^{'}_{BH}}(\dot M_{ADM}) \sim \frac{1}{\dot M_{ADM}}$.

Let us now define {\it the Page point} as the crossing point of both curves i.e. the value of $\dot M_{ADM}$ at which
\begin{equation}
d_{{\cal{A}}_{BH}}(\dot M_{ADM})= d_{{\cal{A}}^{'}_{BH}}(\dot M_{ADM})
\end{equation}
This crossing {\it Page point} corresponds to the type $II_1$ GNS representation i.e. $d_{{\cal{A}}_{BH}}(\dot M_{ADM})=1$ that in this qualitative approach corresponds to $\dot M_{ADM}\sim 1$.

This scheme will reproduce {\it Page's curve} if we associate the curve $d_{{\cal{A}}_{BH}}(\dot M_{ADM})$ with the {\it decrease} of the BH entropy during evaporation and $d_{{\cal{A}}^{'}_{BH}}(\dot M_{ADM})$ with the {\it increase} of the "radiation" entanglement entropy. Both curves cross at the Page point defined above. Thus formally we can define a {\it Murray-von Neumann version of Page curve} defining
\begin{equation}
d_{Page} (\dot M_{ADM}) = min(d_{{\cal{A}}_{BH}}(\dot M_{ADM}),d_{{\cal{A}}^{'}_{BH}}(\dot M_{ADM}) )
\end{equation}
Obviously this suggestion requires much more work to become rigorous, but we believe it captures the algebraic essence of Page's curve and illuminate the role of the type $II_1$ version of BH's.

\subsubsection{A suggestion of  Page's curve for de Sitter}

Following the same {\it clock philosophy} we can think the type $II_1$ factor we associate to dS as the result of transforming the type $II_{\infty}$ dS centralizer by making the extra degree of freedom a real clock. This is what we normally effectively do imposing positivity for the momentum of the extra degree of freedom. However we can do it defining a natural physical clock as discussed in the introduction. To do that we promote the extra degree of freedom coordinate into the inflaton vev $\phi_{inf}$. In the type $II_{\infty}$ version we use a vanishing clock hamiltonian impliying $\dot \phi_{inf}=0$. Hence in this version we define the dS type $II_1$ factor after including a non vanishing $\dot \phi_{inf}$ that we can think as the slow roll parameter $\epsilon$. In that sense the type $II_{\infty}$ can be visualized as a flat inflaton potential with the flow of weights acting as translations of $\phi_{inf}$ in the flat direction. Once we define a clock type $II_1$ factor by some non vanishing $\dot \phi_{inf}$ we can, as we did for the type $II_1$ version of the BH, to define the Murray-von Neumann coupling constant
\begin{equation}
d_{{\cal{A}}_{dS}}(\dot \phi_{inf})
\end{equation}
as a function of $\dot \phi_{inf}$ and where we put in correspondence $\dot M_{ADM}$ for the BH with $\dot \phi_{inf}$ for dS with the inflaton as {\it extra degree of freedom}.

Denoting, as usual in Cosmology $\sqrt{\epsilon}$ the value of $\dot\phi_{inf}$ we could define the dS curves
$d_{{\cal{A}}_{dS}}(\sqrt{\epsilon})$ and $d_{{\cal{A}}_{dS}^{'}}(\sqrt{\epsilon})$. As before we expect for $\epsilon=0$ to have $d_{{\cal{A}}_{dS}^{'}}(\sqrt{\epsilon}=0)=\infty$ and $d_{{\cal{A}}_{dS}}(\sqrt{\epsilon}=0) =0$ i.e. in the no clock , no projection, limit ${\cal{A}}_{dS}$ becomes the type $II_{\infty}$ factor.

Extending naively what we did for the BH we will expect to have the analog of {\it Page point at $\epsilon\sim 1$} that normally we identify, in Cosmology, as the end of inflation and using for the region with $\epsilon>1$ the curve $d_{{\cal{A}}_{dS}^{'}}(\sqrt{\epsilon})$ going to zero for large $\epsilon$. Note that the formal type $II_{\infty}$ limit defined by $d_{{\cal{A}}_{dS}} =0$ that in \cite{Witten2} is interpreted as the no observer case in the second patch is now associated, in this clock picture, with the limit $\epsilon=0$ i.e. with the no clock limit with the flow of weights transforming the value of $\phi_{inf}$ \footnote{The key physics of the type $II_{\infty}$ version of dS is to work, in order to define the action of the commutant, in a Hilbert space $H_{tr}\otimes l^2(N)$ with $H_{tr}$ the GNS Hilbert space defined by the trace. Normally we associate TFD states, for instance the maximal entropy state, to elements in $H_{tr}$ while now we need to work with elements in $l^2(N,H_{tr})$ i.e. with square summable sequences $\Psi(i)$ with $i\in N$ and $\Psi(i)\in H_{tr}$. In this sense the state describing the type $II_{\infty}$ BH can be interpreted, morally speaking, as a square summable sequence of type $II_1$ "maximal entropy" type $II_1$ dS states.}.

\subsection{The type $II_1$ dS factor defined by a squeezed weight}

When we do Cosmology we work with the planar patch, so a natural output of the former discussion is if we can define a natural dS clock on the planar patch. When dealing with the planar patch we will need to deal with a time evolution in the Hilbert space $L^2(\mathbb{R})$ of the extra degree of freedom. The {\it integrable} way to do it will be to consider the functions $|\psi\rangle(t)$ with $|\psi\rangle(t)\in L^2(\mathbb{R})$ to be square integrable i.e. as elements in
\begin{equation}\label{twotimes}
L^2(\mathbb{R},L^2(\mathbb{R})) = L^2(\mathbb{R})\otimes L^2({\mathbb{R}})
\end{equation}
i.e. we will effectively add {\it two extra degrees of freedom}. Physically by introducing an {\it integrable time} we represent the fact that in dS the clock itself is time dependent. The reader can at this point ignore the physics motivation and to proceed with the two copies defined in (\ref{twotimes}). 

\subsubsection{The squeezed automorphism} 
Now instead of using $U(t)$ on $L^2(\mathbb{R})$ we will use a squeezed map $S^t$ defined by the dS Bogolioubov transform for conformal time. In each copy of $L^2(\mathbb{R})$ we have the $t^{obs}$ and $H^{obs}$ coordinate and momentum operators so we can formally define two sets of creation annihilation operators let us say $a_i,a_i^{+}$ for $i=1,2$ and define $S^t$ as the two by two Bogolioubov transformation. 

This allows us to define the {\it squeezed weight} $\omega_s$ such that the corresponding automorphism $\sigma^t_s$ is defined as
\begin{equation}
\sigma^t_s= \sigma^t_{\phi_{dS}}\otimes S^t
\end{equation}
The centralizer of $\omega_s$ will be the suggested definition of the type $II_1$ dS factor relevant for describing Inflationary Cosmology in the planar patch with the corresponding clock Hamiltonian the generator of the squeezed Bogolioubov transform $S^t$. This program has been previously described in \cite{Gomez2,Gomez3}. 

\section{Final remarks and outlook}
The punch line of this note has been the study of the quantum extra degree of freedom needed to implement general covariance  for QFT defined on space-time backgrounds with horizons. We have generically defined clocks whenever this extra quantum degree of freedom is equipped with a positive and hermitian Hamiltonian $H^{clock}$ and we have defined the traces accounting for the expected entropies using a {\it thermal weight}, on the algebra of {\it physical observables} defined by the action of $H^{clock}$, on the Hilbert space of the extra quantum degree of freedom. 

This construction allows us to define a type $II_1$ version of black holes whenever $\dot M_{ADM}$ is non vanishing and to identify a Page curve using the dependence of the type $II_1$  Murray-von Neumann dimension on $\dot M_{ADM}$ and similarly for the dS case once the extra quantum degree of freedom is identified as an inflaton vev.

Many questions remain open and need further clarification. Among them let us mention the following ones:

i) One interesting exercise will be to figure out the way to add an extra degree of freedom to Powers type $III$ factors in order to get a type $II_1$ factor.

ii) To identify a rigorous way to define the {\it clock weights} using a neat definition of good clock states.

iii) To justify the dependence of the Murray-von Neumann coupling constant on $\dot M_{ADM}$ or ( in the case of dS) on $\epsilon$.

iv) To understand how a type $II_1$ involving a non vanishing $\dot M_{ADM}$ ( or a non vanishing $\epsilon$ ) encodes some  non trivial gravitational back reaction quantum gravity effects.

v) How to use hyper finiteness to relate the sort of finiteness of type $II_1$ factors with the quantum gravity finiteness of area formulas for the entropy. In other words we need to deal with the dilemma of the apparent conflict between finiteness of Hilbert space dimension and {\it emergence of time}. Emergence of gravity \cite{Verlinde} is after all more than any other thing {\it time emergency}.

vi) It could be also interesting to go a bit deeper in the potential connection between what we have called {\it integrable time} and {\it complexity}.

vii) Finally it seems that gravity only shows up i.e. emerges, when we link abstract coordinate time with physical {\it clock time}.

\section{Acknowledgments}
This work was supported by grants SEV-2016-0597, FPA2015-65480-P and PGC2018-095976-B-C21. I would like to thank E. Verlinde for an inspiring  conversation in Cambridge. Parts of this work were presented  in the BIRS Meeting {\it "Quantum Information Theory in QFT and Cosmology"}.


\begin{thebibliography}{99}
\bibitem{Bek}J. D. Bekenstein, "Black Holes and the Second Law" Lett. Nuovo Cim.4(1972) 737-740. 
\bibitem{Hawking}
S.~W.~Hawking,
``Particle Creation by Black Holes,''
Commun. Math. Phys. \textbf{43} (1975), 199-220
[erratum: Commun. Math. Phys. \textbf{46} (1976), 206]
 \bibitem{GH}
  G.~W.~Gibbons and S.~W.~Hawking,
  ``Cosmological Event Horizons, Thermodynamics, and Particle Creation,''
  Phys.\ Rev.\ D {\bf 15} (1977) 2738.
  \bibitem{thooft}
  G ’t Hooft, "Dimensional Reduction in Quantum Gravity", Utrecht Preprint THU-93/26, gr-qc/9310006
  \bibitem{Susskind1}
L.~Susskind,
``The World as a hologram,''
J. Math. Phys. \textbf{36} (1995), 6377-6396
doi:10.1063/1.531249
[arXiv:hep-th/9409089 [hep-th]].
\bibitem{LL1}
 S. Leutheusser and H. Liu, "Emergent times in holographic duality" 2112.12156.
 \bibitem{LL2}
 S. Leutheusser and H. Liu, [arXiv:2112.12156 [hep-th]].
\bibitem{Witten1}
E.~Witten,
"Gravity and the Crossed Product''
[arXiv:2112.12828 [hep-th]].
\bibitem{Witten2} 
V.~Chandrasekaran, R.~Longo, G.~Penington and E.~Witten,
``An Algebra of Observables for de Sitter Space,''
[arXiv:2206.10780 [hep-th]].
\bibitem{Gomez1}
C.~Gomez,
``Cosmology as a Crossed Product,''
[arXiv:2207.06704 [hep-th]].
\bibitem{Witten3}
V.~Chandrasekaran, G.~Penington and E.~Witten,
``Large N algebras and generalized entropy,''
[arXiv:2209.10454 [hep-th]].
\bibitem{Witten4}
E.~Witten,
``Algebras, Regions, and Observers,''
[arXiv:2303.02837 [hep-th]].
\bibitem{Witten5}
A.~Strohmaier and E.~Witten,
``The Timelike Tube Theorem in Curved Spacetime,''
[arXiv:2303.16380 [hep-th]].
\bibitem{Corean}
M.~S.~Seo,
``von Neumann algebra description of inflationary cosmology,''
[arXiv:2212.05637 [hep-th]].
\bibitem{Gomez2}
C.~Gomez,
``Entanglement, Observers and Cosmology: a view from von Neumann Algebras,''
[arXiv:2302.14747 [hep-th]].
\bibitem{LL3}
S.~Leutheusser and H.~Liu,
``Subalgebra-subregion duality: emergence of space and time in holography,''
[arXiv:2212.13266 [hep-th]].
\bibitem{Gomez3}
C.~Gomez,
``Clocks, Algebras and Cosmology,''
[arXiv:2304.11845 [hep-th]].
\bibitem{recent1}
K. Jensen, J. Sorce, and A. Speranza, “Generalized entropy for general subregions in quantum gravity,” arXiv:2306.01837 [hep-th].
\bibitem{recent2}
S. Ali Ahmad and R. Jefferson, “Crossed product algebras and generalized entropy for subregions,” arXiv:2306.07323 [hep-th].

\bibitem{recent3}
M.~S.~Klinger and R.~G.~Leigh,
``Crossed Products, Extended Phase Spaces and the Resolution of Entanglement Singularities,''
[arXiv:2306.09314 [hep-th]].
\bibitem{Jones}
V.F.R. Jones "Von Neumann Algebras" Vanderbilt 2015
\bibitem{Gomezcomplex}
C.~G\'omez,
``Complexity and Time,''
Phys. Rev. D \textbf{101} (2020) no.6, 065016
%doi:10.1103/PhysRevD.101.065016
[arXiv:1911.06178 [hep-th]].
\bibitem{Connestaka}
Alain Connes and Masamichi Takesaki,
"The Flow of Weights on Factors of type $III$". Tohoku 
Math Journal 29 (1977) 473
 \bibitem{reference}
S. D. Bartlett, T. Rudolph, and R. W. Spekkens
"Reference frames, superselection rules, and quantum information"
Rev. Mod. Phys. 79, 555 (2007)
\bibitem{AS}
  Y. Aharonov and L. Susskind
  "Charge superselection rule"
Phys. Rev. 155, 1428 (1967)
\bibitem{Susskind2}
L.~Susskind,
``A Paradox and its Resolution Illustrate Principles of de Sitter Holography,''
[arXiv:2304.00589 [hep-th]].

\bibitem{Torroba1} 
R. Bousso, "Bekenstein Bounds in de Sitter and Flat Space," JHEP04(2001) 035,hep-th/0010252.
\bibitem{Torroba2}T. Banks, "More Thoughts on the Quantum Theory of Stable de Sitter Space" arXiv:hep-th/0503066.
\bibitem{Torroba3} T. Banks, B. Fiol, and A. Morisse, "Towards a Quantum Theory of de Sitter Space" arXiv:hep-th/0609062.
\bibitem{Torroba4}  X. Dong, E. Silverstein, and G. Torroba, "De Sitter Holography and Entanglement Entropy" arXiv:1804.08623.
\bibitem{Page}
D.~N.~Page,
``Information in black hole radiation,''
Phys. Rev. Lett. \textbf{71} (1993), 3743-3746
doi:10.1103/PhysRevLett.71.3743
[arXiv:hep-th/9306083 [hep-th]].
\bibitem{wittenmedal}
E.~Witten,
``APS Medal for Exceptional Achievement in Research: Invited article on entanglement properties of quantum field theory,''
Rev. Mod. Phys. \textbf{90} (2018) no.4, 045003
\bibitem{Susskind3}
H.~Lin and L.~Susskind,
``Infinite Temperature's Not So Hot,''
[arXiv:2206.01083 [hep-th]].
\bibitem{ABc}
Y.~Aharonov and D.~Bohm,
``Time in the Quantum Theory and the Uncertainty Relation for Time and Energy,''
Phys. Rev. \textbf{122} (1961) no.5, 1649-1658
\bibitem{Verlinde}
E.~P.~Verlinde,
``On the Origin of Gravity and the Laws of Newton,''
JHEP \textbf{04} (2011), 029
%doi:10.1007/JHEP04(2011)029
[arXiv:1001.0785 [hep-th]].


\end{thebibliography}
\end{document}